\documentclass[twocolumn]{aastex62}

\usepackage{bm}

\newcommand{\pr}[1]{\ensuremath{\left(#1\right)} }
\newcommand{\pd}[2]{\ensuremath{\frac{\partial #1}{\partial #2} }}

\newcommand{\eqref}[1]{Equation (\ref{#1})}

\newcommand{\K}{\ensuremath{\rm \,K }}
\newcommand{\AU}{\ensuremath{\rm \,AU }}

\newcommand{\etaO}{\eta_{\rm O} }
\newcommand{\etaH}{\eta_{\rm H} }
\newcommand{\etaA}{\eta_{\rm A} }

\newcommand{\Am}{{\rm Am} }

\newcommand{\mum}{$\rm \mu m$}

\renewcommand{\d}{{\rm d}}
\newcommand{\dx}{{\rm d}x}
\newcommand{\dy}{{\rm d}y}
\newcommand{\dz}{{\rm d}z}
\newcommand{\dV}{{\rm d}V}
\newcommand{\dr}{{\rm d}r}

\newcommand{\dmu}{{\rm d}\mu}
\newcommand{\Jth}{J_{\rm th}}
\newcommand{\Hth}{H_{\rm th}}
\newcommand{\Kth}{K_{\rm th}}

\newcommand{\vvec}[1]{ \pr{ \begin{array}{c} #1 \end{array} } }
\newcommand{\vK}{v_{\rm K}}
\newcommand{\cs}{c_{\rm s}}
\newcommand{\zb}{z_{\rm b}}

\defcitealias{Okuzumi2015The-Nonlinear-O}{OI15}
\defcitealias{Mori2016Electron-Heatin}{MO16}

\begin{document}

\title{
Temperature Structure in the Inner Regions of Protoplanetary Disks: Inefficient Accretion Heating Controlled by Nonideal Magnetohydrodynamics
}

\shortauthors{Mori et al.}
\def\myemail{mori.s@geo.titech.ac.jp}
\def\titech{Department of Earth and Planetary Sciences, Tokyo Institute of Technology, Meguro-ku, Tokyo, 152-8551, Japan}
\def\tsinghua{Institute for Advanced Study and Tsinghua Center for Astrophysics, Tsinghua University, Beijing 100084, China}

\author[0000-0002-7002-939X]{Shoji Mori}       \affiliation{\titech; \href{mailto:\myemail}{\rm \myemail}}
\author[0000-0001-6906-9549]{Xue-Ning Bai}     \affiliation{\tsinghua; \href{mailto:xbai@tsinghua.edu.cn}{\rm xbai@tsinghua.edu.cn}}  
\author[0000-0002-1886-0880]{Satoshi Okuzumi}  \affiliation{\titech}



\begin{abstract}
The gas temperature in protoplanetary disks (PPDs) is determined by a combination of irradiation heating and accretion heating, with the latter conventionally attributed to turbulent dissipation. However, recent studies have suggested that the inner disk (a few au) is largely laminar, with accretion primarily driven by magnetized disk winds, as a result of nonideal magnetohydrodynamic (MHD) effects from weakly ionized gas, suggesting an alternative heating mechanism by Joule dissipation. We perform local stratified MHD simulations including all three nonideal MHD effects (ohmic, Hall, and ambipolar diffusion) and investigate the role of Joule heating and the resulting disk vertical temperature profiles. We find that in the inner disk, as ohmic and ambipolar diffusion strongly suppress electrical current around the midplane, Joule heating primarily occurs at several scale heights above the midplane, making the midplane temperature much lower than that with the conventional viscous heating model. Including the Hall effect, Joule heating is enhanced/reduced when the magnetic fields threading the disks are aligned/anti-aligned with the disk rotation, but it is overall ineffective. Our results further suggest that the midplane temperature in the inner PPDs is almost entirely determined by irradiation heating, unless viscous heating can trigger thermal ionization in the disk innermost region to self-sustain magnetorotational instability turbulence.
\end{abstract}%


\section{Introduction}

The temperature structure of protoplanetary disks is essential for understanding many processes of planet
formation. Particularly relevant is dust composition.
Outside the snow line where water condenses into ice, the dust is mainly composed of water ice and
silicate \citep{Lodders2003Solar-System-Ab}. The icy dust aggregates are more sticky
\citep{Wada2009Collisional-Gro} and are likely to directly grow to planetesimals via collisional sticking
\citep{Okuzumi2012Rapid-Coagulati,Kataoka2013Fluffy-dust-for}, making the initial stage of planet formation
proceed differently inside and outside of the snow line.
Disk temperature structure is also important for understanding the water content of solar system bodies,
since it directly reflects the water content of the accreted material, which is temperature sensitive.
For instance, the Earth's ocean is only 0.023 wt\% of the total Earth mass, whereas the water content of
comets can be as high as 50 wt\% \citep[e.g.,][]{AHearn2011EPOXI-at-Comet-}. Similar low water content
($<5$ wt\%) is inferred from the TRAPPIST-1 system \citep{Grimm2018The-nature-of-t}.  

The disk temperature is determined mainly by two heating mechanisms: irradiation and accretion heating.
Irradiation from the central star directly heats the surface and determines the bulk disk temperature \citep[e.g.,][]{Kusaka1970Growth-of-Solid,Chiang1997Spectral-Energy}, 
and it generally results in a vertical
temperature profile that peaks at disk surface.
Accretion heating is conventionally considered to be due to viscous dissipation mediated by turbulence, a process
that also drives disk accretion.
It is commonly described by the Shakura-Sunyaev $\alpha$-disk model \citep{Shakura1973Black-holes-in-}, 
with effective viscosity $\nu$ expressed as
\begin{equation}
	\nu = \alpha c_{s} H \ ,
\end{equation}
where $c_{s}$ is the sound speed, $H = \cs/\Omega$ is the gas scale height ($\Omega$ is the
Keplerian angular velocity), and the strength of viscosity/turbulence is characterized by the dimensionless
parameter $\alpha$. The heating rate and the viscously-driven accretion rate are then proportional to $\alpha$
and density. With constant $\alpha$, heating concentrates in the disk midplane, and makes disk temperature peak
at the midplane.
For typical PPD accretion rate of $\sim10^{-8}{M_\sun}$ yr$^{-1}$ and a conventional disk model (e.g.,
minimum-mass solar nebula; \citealp{Weidenschilling1977Aerodynamics-of,Hayashi1981Structure-of-th}), it can be found that viscous heating dominates only at sub-AU scale,
beyond which the disk temperature is mainly determined by irradiation.
With higher accretion rate in early phases of disk evolution, viscous heating likely dominates to larger
distances, even up to a few 10s of AU during accretion outbursts \citep{Cieza2016Imaging-the-wat}.

Turbulence in protoplanetary disks is thought to be generated mainly by the magnetorotational
instability \citep[MRI;][]{Balbus1991A-powerful-loca}. It occurs when the magnetic field coupled with the
ionized gas is stretched by shear and rotation, and can generate strong magnetic turbulence with
$\alpha \sim 10^{-3}$--$10^{-2}$ when the gas is well ionized
\citep[e.g.,][]{Hirose2006Vertical-Struct,Flaig2010Vertical-struct}.
However, PPDs are extremely weakly ionized, and the coupling between gas and magnetic fields is substantially
weakened by three non-ideal MHD effects, i.e. Ohmic diffusion, the Hall effect, and ambipolar diffusion
\citep[e.g.,][]{Wardle2007Magnetic-fields,Bai2011Magnetorotation}.
Ohmic diffusion tends to dominate in regions where the gas density is high and the magnetic field is weak. Ambipolar diffusion
tends to be important in low-density regions and relatively strong magnetic field. The Hall-dominated regime lies in
between. In the inner disks, it turns out that Ohmic, Hall and ambipolar diffusion dominates in the midplane,
intermediate and surface layers, respectively, and all these effects strongly affect the properties of the MRI
\citep[e.g.,][]{Jin1996Damping-of-the-,Wardle1999The-Balbus-Hawl,Balbus2001Linear-Analysis,Desch2004Linear-Analysis,Kunz2004Ambipolar-diffu}.

Combined with the ionization conditions in PPDs, it is well known that in the inner PPD,
Ohmic resistivity stabilizes the MRI around the midplane
\citep[e.g.,][]{Gammie1996Layered-Accreti,Sano2000Magnetorotation,Fleming2003Local-Magnetohy,Dzyurkevich2013Magnetized-Accr}.
Without including other non-ideal MHD
effects, the vigorous MRI turbulence is present only in the surface layer, leading to the picture of layered accretion.
The resulting vertical temperature profile was investigated by \citet{Hirose2011Heating-and-Coo}, who found lower
midplane temperatures than viscous models with a constant $\alpha$ parameter.
This is because the heating by turbulent dissipation peaks at disk surface and is lost more directly by radiation cooling
instead of heating the midplane.

In the recent years, however, it has been realized that ambipolar diffusion can stabilize the MRI in the upper layer of inner PPDs
\citep[e.g.,][]{Bai2011Effect-of-Ambip,Gressel2015Global-Simulati,Bai2013aWind-driven-Acc,Bai2013bWind-driven-Acc}.
With MRI fully suppressed, disk accretion and evolution is driven by magnetized disk wind \citep[e.g.,][]{Bai2017Global-Simulati}.
whereas the non-dissipative Hall effect has more subtle behaviors that depend on the polarity of the net vertical
field threading the disk 
\citep[e.g.,][]{Sano2002aThe-Effect-of-t,Sano2002bThe-Effect-of-t,Kunz2008On-the-linear-s,Lesur2014Thanatology-in-,Bai2014Hall-effect-Con,Bai2015Hall-Effect-Con,Tsukamoto2015Bimodality-of-C,Simon2015Magnetically-dr,Bai2017Hall-effect-Med,Bai2017Global-Simulati}.

In a largely laminar disk, 
accretion heating profile is then primarily determined by magnetic diffusivity instead of the turbulent viscosity.
This heating mechanism is fundamentally different from viscous dissipation: there is no simple relation between
heating rate and wind-driven accretion rate. The heating rate is merely related to the vertical profile of
magnetic diffusivities and electric current. Therefore, detailed disk microphysics is essential to properly
calculate the Joule heating rate. Furthermore, the presence of the Hall effect can amplify or reduce
horizontal magnetic field depending on polarity 
\citep{Lesur2014Thanatology-in-,Bai2014Hall-effect-Con,Bai2015Hall-Effect-Con,Simon2015Magnetically-dr}, and is expected to yield different temperature profiles.

In this paper, we study the rate of Joule heating in the inner PPDs by means of local non-ideal MHD simulations that
incorporate all three non-ideal MHD effects, calculate the resulting vertical temperature profiles, and discuss their
physical implications on planet formation.  The plan of this paper is as follows.
In Section \ref{sec:methods}, we describe our simulation setup and model parameters.
In Section \ref{sec:fid}, we present the results of our simulations for a fiducial set of parameters,
focusing on the energy dissipation and temperature profiles.  
A parameter study is presented in Section \ref{sec:ps}, investigating when and how the accretion heating is inefficient.
Limitations of our local simulations and implications on the planet formation are discussed in Section \ref{sec:discuss}
before we summarize in Section \ref{sec:Summary}.

\section{Methods and Model}\label{sec:methods}

\subsection{Numerical Method}\label{ssec:num-meth}

We perform MHD simulations in a local shearing box \citep{Goldreich1965I.-Gravitationa,Hawley1995Local-Three-dim} using Athena
\citep{Stone2008Athena:-A-New-C}, an open source MHD code based on the Godunov method
with constrained transport to preserve the divergence-free condition of magnetic fields.	
A shearing box is centered on a fixed radius $R_0$ and works in a frame that is corotating
with its Kepler angular velocity $\bm{\Omega}$. By ignoring disk curvature, one employs
cartesian coordinates ($x$, $y$, $z$) for the radial, azimuthal and vertical dimensions.
The orbital advection scheme \citep{Masset2000FARGO:-A-fast-e} described in \citet{Stone2010Implementation-} is used, where the velocity is decomposed into the
background Kepler velocity $\bm{v}_{\rm K}= -(3/2)\Omega x \hat{\bm{y}}$ and 
deviation from  it, $\bm{v}$.
We take into account all three non-ideal MHD effects: Ohmic diffusion, the Hall effect,
and ambipolar diffusion,
which are characterized by diffusion coefficients $\etaO, \etaH,$ and $\etaA$, 
respectively. We solve the following basic equations,
\begin{equation}
	\pd{\rho}{t} + \vK \pd{\rho}{y} +  \nabla \cdot \pr{\rho \bm{v} } = 0  \label{basicEqDensity},	    
\end{equation}
\begin{eqnarray}
&&\pd{\rho \bm{v}} {t} + \vK \pd{\rho\bm{v}}{y} + \nabla \cdot \pr{\rho \bm{v}\bm{v} - \frac{\bm{BB}}{4 \pi} + \pr{P + \frac{B^{2}}{8\pi} } \mathsf{ I }   } \nonumber \\
	&&\qquad= 2 \Omega \rho v_{y} \hat{\bm{x}} - \frac{1}{2} \Omega \rho v_{x} \hat{\bm{y}} - \rho\Omega^{2} \bm{z},   
\end{eqnarray}
\begin{equation}
\pd{\bm{B}}{t} = \nabla \times ((\bm{v}_{\rm K}+{\bm v})\times \bm{B} - c{\bm E}') 
	\label{basicEqMagnet} , 
\end{equation}
where 
$\rho$ is the gas density, 
$\bm{B}$ is the magnetic field, 
$\mathsf{ I } $ is the identity tensor,
$P$ is the gas pressure, 
$c$ is the speed of light,
and $\bm{E}'$ is the electric field as measured in the frame comoving with the neutral gas. The generalized Ohm's law relates ${\bm E'}$ to the current density 
$\bm{J} = (c/4\pi)\nabla \times \bm{B}$ as
\begin{equation}
\bm{E}' = \frac{4\pi}{c^2}(\etaO \bm{J}  + \etaH \bm{J}\times \hat{\bm{B}} + \etaA \bm{J}_{\perp}),
\label{eq:E'}
\end{equation} 
where $\hat{\bm{B}}$ is the unit vector along the magnetic field and
$\bm{J}_{\perp} = - (\bm{J} \times \hat{\bm{B}} )\times \hat{\bm{B}}$
is the current density perpendicular to the magnetic field.
The sign of the vertical magnetic field is taken to be positive when its direction is the same as $\bm{\Omega}$, which is along $\bm{z}$. Note that the disk is vertically
stratified by including vertical gravity $\propto-\Omega^2z$.
An isothermal equation of state is adopted, with $P=\rho c_s^2$ and $c_s$ is the isothermal sound
speed. Length scales are then measured in disk scale height $H\equiv c_s/\Omega$.
We impose the shearing periodic boundary condition for $x$, the periodic boundary condition for $y$,
and the outflow boundary condition for $z$.

Magnetic diffusivities depend on the number densities of all charge carriers.
We calculate the number densities of electrons, ions and charged grains in the disk interior, by
considering ionization by cosmic rays (CR), stellar X-rays, and short-lived radionuclides, and their
recombination in the gas and on dust grain's surface. 
The ionization prescriptions are the same as those adopted in \citet{Bai2011Magnetorotation}, where 
we adopt CR ionization rate profile described in \citet{Sano2000Magnetorotation} which is based on \citet{Umebayashi1981Fluxes-of-Energ},
a fitting formula in \citet{Bai2009Heat-and-Dust-i} for X-ray ionization with X-ray luminosity
$L_X=10^{30}$ ergs s$^{-1}$ and X-ray temperature $T_X=5$ keV. Ionization rate by radionuclides
is taken to be constant at $7.6\times10^{-19}$ s$^{-1}$.
For ionization and recombination reactions, we use the same model as used in
\citet{Mori2016Electron-Heatin}.
We represent all ion species with a single species by following \citet{Okuzumi2009Electric-Chargi}. 
The diffusion coefficients are expressed in terms of the Hall parameters of individual charged species
as is commonly done \citep[e.g.,][]{Wardle2007Magnetic-fields,Bai2011Magnetorotation}.
In regimes of interest (when small dust grains are scarce), the diffusion coefficients $\etaO, \etaH,$
and $\etaA$ can be found to be proportional to $B^{0}$, $B^{1}$, and $B^{2}$ respectively,
with $f_{\rm dg} (a/1\mu {\rm m})^{-2} < 1$
\citep{Xu2016On-the-Grain-mo}, where $f_{\rm dg}$ and $a$ are the dust-to-gas mass ratio and radius of
small dust grains, respectively. This condition is generally satisfied in PPDs with grain growth, and 
is marginally satisfied in the simulations presented in this study.
A diffusivity table is then obtained by fixing grain size, dust abundance, gas temperature,
expressing $\eta_O$, $Q_{\rm H}\equiv\eta_H/B$ and $Q_{\rm A}\equiv \eta_A/B^{2}$, as a function of gas
density and ionization rate.

We also take into account the ionization by far-ultraviolet radiation (FUV) in the surface layer,
where FUV can substantially enhance the level of ionization \citep{Perez-Becker2011Surface-Layer-A}.
Similar to the treatment of \citet{Bai2013aWind-driven-Acc}, we impose an ionization fraction of
$3\times 10^{-5}$ in the FUV layer (from which magnetic diffusivities can be calculated), with a
penetration depth of 0.03 g cm$^{-2}$ to the vertical boundary. A smooth transition is the imposed
over a few grid cells to join the magnetic diffusivity of the bulk disk.

The importance of non-ideal MHD effects is characterized by Elsasser numbers, which read
\begin{eqnarray}
	\Lambda = \frac{v_{\rm A}^{2}}{\etaO\Omega} \ , \quad
	\chi =   \frac{v_{\rm A}^{2}}{\etaH\Omega} \ , \quad
	\Am = \frac{v_{\rm A}^{2}}{\etaA\Omega} \ ,
\end{eqnarray}
where $v_{\rm A}=B/\sqrt{4 \pi \rho}$ is the Alfv\'en speed. Non-ideal MHD effects are
considered strong when the Elsasser numbers are around or below unity.

A mass outflow from vertical boundaries is naturally produced in our simulations that can reduce
the total mass in the simulation box. To facilitate our analysis, 
we add mass to the system at each time step to keep this total mass unchanged to achieve steady
state over long timescales. This treatment has a negligible effect on the overall dynamics.

Finally, although we aim to study the vertical temperature in disks, we still assume an
isothermal equation of state in the simulations for simplicity. This is because ionization-recombination
chemistry typically depends weakly on disk temperature, allowing us to reconstruct the temperature
profile from the energy dissipation profile in an isothermal simulation, expecting that the dynamics is
not to be strongly affected under the updated temperature profiles. In the mean time,
we also perform the simulations with different isothermal temperatures in Section \ref{ssec:disc-temp}
to assess the validity of this approach.

	\subsection{Simulation Setup}\label{ssec:setting}

\begin{deluxetable*}{lll}
\tablecaption{Summary of the parameters \label{tab:para}}
\tablehead{ \colhead{Parameter} & \colhead{Values} & \colhead{Description}  }
\startdata
$r$ [AU] 					&[0.2, 0.5, 1\tablenotemark{*}, 2 ] & Distance from the central star \\
$\Sigma_{0}$ [g cm$^{-2}$]	 		&[170, 1700\tablenotemark{*}, 17000] & Surface density at 1 AU \\
$f_{\rm dg} $				&[$10^{-3}$, $10^{-4}$\tablenotemark{*},$10^{-5}$] & Dust-to-gas mass ratio \\
$\beta_{0}$ 			&[10$^{3}$, 10$^{4}$, 10$^{5}$\tablenotemark{*}, 10$^{6}$] & Initial gas-to-magnetic pressure ratio at the midplane\\
${\rm sgn}\pr{ B_{z} }$	&[$+1$\tablenotemark{*}, $-1$] & Alignment of the initial magnetic field with the disk rotation axis \\
\enddata
\tablenotetext{*}{Fiducial value} 
\end{deluxetable*}

Following \citet{Bai2013bWind-driven-Acc} and \citet{Bai2014Hall-effect-Con}, all our simulations are quasi-1D by using a
computational domain size of ($L_{x}$, $L_{y}$, $L_{z}$) = ($0.48H, 0.48H ,16 H$) with a computational
grid of $4 \times 4 \times 192$ cells. This is because the flow in the inner regions is expected to be
largely laminar.

The initial gas density profile is taken to be a Gaussian $\rho = \rho_{0} {\rm exp}(-z^{2}/(2H^{2}) )$, 
where $\rho_{0}$ is the initial gas density at the midplane, with initial perturbations.
The amplitudes of the initial density perturbations and velocity perturbations are 1\% and 0.4\% of the
background values, respectively. We set a density floor of $10^{-8} \rho_{0}$.

The initial magnetic field configuration is given by the sum of a uniform vertical field and sinusoidal components,
$\bm{B}_{0} = (0,B_{0}/\sqrt{2} \sin(\pi z/L_{z}), B_{0} )$.
The background field strength $B_{0}$ is characterized by the midplane plasma beta
$\beta_{0}= 8 \pi \rho_{0} c_{\rm s}^{2}/B_{0}^{2}$ (the ratio of the gas pressure
$P_{0}= \rho_{0} c_{\rm s}^{2}$ to the magnetic pressure at the midplane). The sign of this background
vertical field can be either positive or negative, whose dynamics will be different due to the Hall effect.
The sinusoidal toroidal field is included to help poloidal field grow into a physical field geometry (discussed in Section \ref{ssec:disc-sym}), 
where the field lines at upper and lower disk surface are in the opposite directions.
We have confirmed that the sinusoidal component does not affect the final state.

In code units, we adopt $H=c_s=\Omega=1$, and $\rho_0=1$. For magnetic field, factors of
$\sqrt{4\pi}$ are further absorbed so that magnetic pressure is simply given by $B^2/2$
(as opposed to the equations we have written which are in cgs/Gauss units).

The magnetic diffusivities in the midplane region can become excessively large due to the extremely weak
level of ionization, causing excessively small simulation timesteps. To alleviate the issue, we impose a
diffusivity cap $\eta_{\rm cap}$ so that when the sum of all diffusivity coefficients exceeds the cap,
each of them is reduced by the same factor so that its sum just reaches $\eta_{\rm cap}$.
In practice, we choose $\eta_{\rm cap}=200\cs H$. Note that this is much larger than the value of
$10\cs H$ that is more commonly adopted \citep[e.g.,][]{Bai2013aWind-driven-Acc,Gressel2015Global-Simulati}. Here, we choose a higher value for $\eta_{\rm cap}$ because we have found that convergence of main diagnostic quantities for our purposes (e.g., work done by shear) converge for $\eta_{\rm cap} \geq 100\cs H$.

\subsection{Simulation Parameters}
The parameters adopted in our simulations are summarized in Table \ref{tab:para}.
Our fiducial model assumes the minimum-mass solar nebula (MMSN) model
\citep{Weidenschilling1977Aerodynamics-of,Hayashi1981Structure-of-th} at 1 AU containing 0.1 \mum-sized
dust grains with the dust-to-gas ratio $f_{\rm dg}$ of $10^{-4}$ and the initial midplane plasma beta $\beta_{0}$ of $10^{5}$.
We take the fiducial value of $f_{\rm dg}$ to be lower than the interstellar value $\approx 0.01$ considering the situation where most submicron-sized grains have already been incorporated into larger solid particles/bodies. As shown by \citet{Ormel2013The-Fate-of-Pla}, a reduction of $f_{\rm dg}$ by a factor of $10^2$--$10^3$ from the interstellar value best represents this situation \citep[see also][]{Birnstiel2011Dust-size-distr}.
The surface density profile is given by $\Sigma=1700(r/\AU)^{-3/2}~\rm g~cm^{-2}$.
The temperature profile of the bulk disk is set to $T = 110 (r/\AU)^{-3/7}~\K$, following \citet{Chiang1997Spectral-Energy} corresponding to a disk temperature set by stellar irradiation (the model is described in Section \ref{ssec:res-temp}).
This fiducial parameter set will produce the accretion rate of $\sim10^{-8}M_\sun~\rm yr^{-1}$ \citep[e.g.,][]{Bai2014Hall-effect-Con}, which corresponds to PPDs around typical T Tauri stars.

The parameter sets represent inner regions ($r=$0.2--5 AU) of young PPDs with a grain abundance corresponding
to some level of grain growth.
We vary the initial midplane plasma beta $\beta_{0}$ from 10$^{3}$ to 10$^{6}$, with the sign of net vertical field
taken to be either positive or negative.

\subsection{Energy Equations}\label{ssec:energyeq}

Upon achieving a steady state in our isothermal simulations, we will need to analyze the energy transport
in the system.
In doing so, we separate total energy density $E_{\rm tot}$ into internal energy density $E_{\rm int}$,
and mechanical energy density $E_{\rm mec} = \rho \bm{v}^{2}/2 +  \rho \Omega^{2}z^{2}/2 + B^{2}/(8\pi)$.
Note that here the velocity $\bm{v}$ already has Keplerian shear subtracted. As we adopt an isothermal
equation of state where $E_{\rm int}$ does not enter self-consistently, we first
focus on the equation for mechanical energy to discuss its energy balance and dissipation profiles, and
then we discuss how we reconstruct the more realistic temperature profiles from the post-process simulation data.

\subsubsection{Dissipation Profiles}

In a shearing-box, the equation of mechanical energy conservation is given by
\citep[e.g.,][]{Balbus1998Instability-tur, Stone2010Implementation-}
\begin{eqnarray}\label{eq:mech-energy-rate}
	&&\pd{ E_{\rm mec} }{t} + \vK \pd{E_{\rm mec} }{y} + \nabla \cdot \bm{F}_{\rm mec} \nonumber\\
	&&=  -q_{\rm Joule} + P \nabla \cdot \bm{v} + w_{\rm str}
\end{eqnarray}
where 
\begin{eqnarray}\label{eq:mech-energy-flux}
	&&\bm{F}_{\rm mec}  = \bm{v} \pr{ \frac{1}{2}\rho v^{2} +  P + \frac{1}{2} \rho \Omega^{2} z^{2} }  \nonumber \\
	&&\hspace{3.2em} + \frac{B^{2}\bm{v}}{4\pi} - \frac{(\bm{v}\cdot\bm{B})\bm{B}}{4\pi}  + \frac{c}{4 \pi} \bm{E}' \times \bm{B} \\
	&&q_{\rm Joule}\equiv \bm{J} \cdot \bm{E}'/c\ ,\qquad
	w_{\rm str} \equiv \frac{3}{2}  \Omega \pr{\rho v_{x} v_{y} - \frac{B_{x}B_{y}}{4 \pi} } \ .
\end{eqnarray}
Here, $\bm{F}_{\rm mec}$ is the energy flux of mechanical energy, which consists of hydrodynamic (first three) and magnetic (last three) terms, respectively.
The last term of $\bm{F}_{\rm mec}$, 
$(c/4\pi)\bm{E}'\times\bm{B}=(\etaO+\etaA)\bm{J}\times\bm{B}-\etaH B\bm{J}_{\perp}$,
 represents the Poynting flux of non-ideal MHD effects.
The term $P \nabla \cdot \bm{v}$ represents the mechanical work ($P\d V$ work) done on the fluid.
The term $q_{\rm Joule}$ represents irreversible energy dissipation by Joule heating, 
with the minus sign before $q_{\rm Joule}$ in \eqref{eq:mech-energy-flux} meaning that this dissipation comes at the cost of mechanical energy.
Of the three non-ideal MHD effects, only Ohmic and ambipolar diffusion generate heat, whereas the Hall effect is dissipationless: substituting \eqref{eq:E'} into $q_{\rm Joule}$ gives $q_{\rm Joule} = (4 \pi/c^{2}) ( \etaO \bm{J}^{2}  + \etaA \bm{J}_{\perp}^{2} )$.
The term $w_{\rm str}$ represents the work done by the Reynolds and Maxwell stresses
through shear,
which injects mechanical energy into the system.

In our simulations, there is energy loss through the vertical boundary through a disk wind.
Globally, the energy balance thus involves energy injection by $w_{\rm str}$, which is then
consumed by 1). Joule dissipation and 2). energy loss through $P\d V$ work and disk outflow.
We emphasize that in the conventional scenario of viscously-driven accretion, the injected energy
$w_{\rm str}$ (i.e., now being the viscous stress) is locally dissipated. However, this no longer holds
in the case of wind-driven accretion. A more detailed discussion is presented in Appendix \ref{sec:cons}.

\subsubsection{Temperature Profiles}\label{ssec:temp}

We here use the energy dissipation profile obtained from the simulation to estimate the temperature
profile. \citet{Hubeny1990Vertical-struct} derived the analytical formula of the temperature profile
in the accretion disk with a viscosity profile. 
We here extend the formula by taking into account of the following points. 

First, \citet{Hubeny1990Vertical-struct} assumed that the dissipated energy equals
to the work done by the viscous stress.
However, the rate profiles of the injected and dissipated energy are different, as we have discussed
above. Second, heating by stellar irradiation is known to be important in PPDs, with heating rate denoted by $q_{\rm irr}$.
Thus, we solve radiative transfer assuming that 
the net heating rate per unit volume, $q$, is given by the sum
$q\equiv q_{\rm Joule}+q_{\rm irr}$. 
In addition, the dissipation profile $q_{\rm Joule}$ can be asymmetric about the disk
midplane, thus we do not assume reflection symmetry of the $q$ profile about the midplane as used in
\citet{Hubeny1990Vertical-struct}. 

The derivation and further discussion are described in Appendix \ref{sec:app}, and the resulting 
temperature profile is given by
\begin{equation}\label{eq:T}
	T (z) = \pr{ \frac{3\mathcal{F}_{+\infty} }{4\sigma} }^{1/4} \pr{\tau_{\rm eff}  
					+    \frac{  1 }{\sqrt{3}}  + \frac{q }{3  \rho \kappa_{\rm R} \mathcal{F}_{+\infty} }  } ^{1/4} \ ,
\end{equation}
where
\begin{eqnarray}
	\tau_{\rm eff}(z) &=& \frac{1}{\mathcal{F}_{+\infty}} \int_{z}^{+\infty} \rho\kappa_{\rm R}  \mathcal{F}(z) \dz'
\nonumber \\ 
&=& \int_{z}^{+\infty} \rho\kappa_{\rm R}  \pr{1 - \frac{1}{\mathcal{F}_{+\infty} }\int_{z'}^{+\infty} q \dz'' } \dz'  \ , \label{eq:taueff}
\end{eqnarray}
\begin{equation}  
	\mathcal{F}_{+\infty} =  \frac{1}{2/\sqrt{3}+  \tau_{R, \rm tot} }\pr{ \frac{\Gamma}{\sqrt{3}} +\int_{-\infty}^{+\infty} \rho\kappa_{\rm R}  \pr{\int_{z}^{+\infty} q \dz'} \dz}   \,  \label{eq:Finf}
\end{equation}
\begin{equation}
	\tau_{R, \rm tot} = \int_{-\infty}^{+\infty} \rho\kappa_{\rm R}  \dz \,
\end{equation}
\begin{equation}
	\Gamma = \int_{-\infty}^{+\infty} q \dz \ .
\end{equation}
Here, $\kappa_{\rm R}$ is the Rosseland mean opacity,
$\mathcal{F}_{+\infty}$ is the radiative flux at $z=+\infty$, 
$\Gamma$ is the total heating energy rate, 
and $\sigma$ is the Stefan-Boltzmann constant.
We have also assumed that the scattering coefficient is much smaller than the absorption coefficient.
The effective optical depth $\tau_{\rm eff}$ represents the radiative-flux-weighted optical depth. Its contribution to disk temperature illustrates how disk temperature is enhanced by heat accumulation in the disk,
which is crucial for understanding how accretion heating increases disk temperature.

For simplicity, throughout this work, we assume constant opacity of $ \kappa_{\rm R} = 5 (f_{\rm dg}/0.01) ~{\rm cm^{2}~g^{-1}}$, which is sufficient for the demonstrative purpose on the discussion of disk heating mechanisms.

To compare the temperature profiles from our simulations with those from the conventional models, 
we consider the two different heating models: an ``equivalent" viscous model assuming local energy dissipation, and a conventional constant-$\alpha$ viscous model. 

For the equivalent viscous model, the heating profile is given by 
\begin{equation}\label{eq:Qvis-loc}
	q_{\rm vis} (z) = -\frac{3\Omega}{2}\frac{B_{x}B_{y}}{4 \pi} \ .
\end{equation}
Note that only Maxwell stress is included because
in a laminar disk wind, the Reynolds (hydrodynamic) stress is generally negligible and is also unrelated with dissipation. 
This model corresponds to the case where the work done by Maxwell stress is locally dissipated as if the system is turbulent. 
Comparison with this model will show the importance of heating profile in controlling disk temperature. 

For the conventional constant-$\alpha$ model, viscosity is taken to be vertically constant given by $\nu=\alpha \cs H$,
with viscous heating rate
\begin{equation}\label{eq:Qvis-alpha}
	q_{\rm vis} (z)= \frac{9}{4} \alpha \rho \cs^{2} \Omega \ .
\end{equation}
The value of $\alpha$ is set by requiring that the resulting steady-state mass accretion rate, $\dot{M}=3\pi\alpha \Sigma \cs H$, matches the mass accretion rate estimated from the simulations \citep{Bai2013aWind-driven-Acc}:
\begin{equation}\label{eq:maccr}
	\dot{M}=\frac{2 \pi}{\Omega}\int^{\zb}_{-\zb}  T_{xy}  \dz + \frac{8\pi}{\Omega}r \left| T_{zy} \right|_{\zb}\ ,
\end{equation}
where $ T_{xy} = \rho v_{x} v_{y} - B_{x} B_{y}/4\pi$ and  $T_{zy} = - B_{z} B_{y}/4\pi$, corresponding to contributions from radial and vertical (wind) transport of angular momentum.
Here, we simply take the height of the base of the wind to be 4 $H$, $\zb$ (which is close to values obtained more systematically, e.g., \citealp{Bai2014Hall-effect-Con}).  

\section{Fiducial Run}\label{sec:fid}

We begin by discussing the results of the fiducial run. The outcome of the simulations is
very similar to those presented in \citet{Bai2013aWind-driven-Acc} and \citet{Bai2014Hall-effect-Con}, where the
system relaxes to a laminar state over a few tens of orbits.

We first briefly discuss the overall properties of the gas dynamics and magnetic
field profiles in Section \ref{ssec:profs}, focusing on dissipation by Joule heating.
We then discuss the temperature profiles resulting from
Joule heating and irradiation.

\subsection{Gas Dynamics and Dissipation}\label{ssec:profs}



\begin{figure*}[t] 
	\centering
	\includegraphics[width=0.49\hsize,clip]{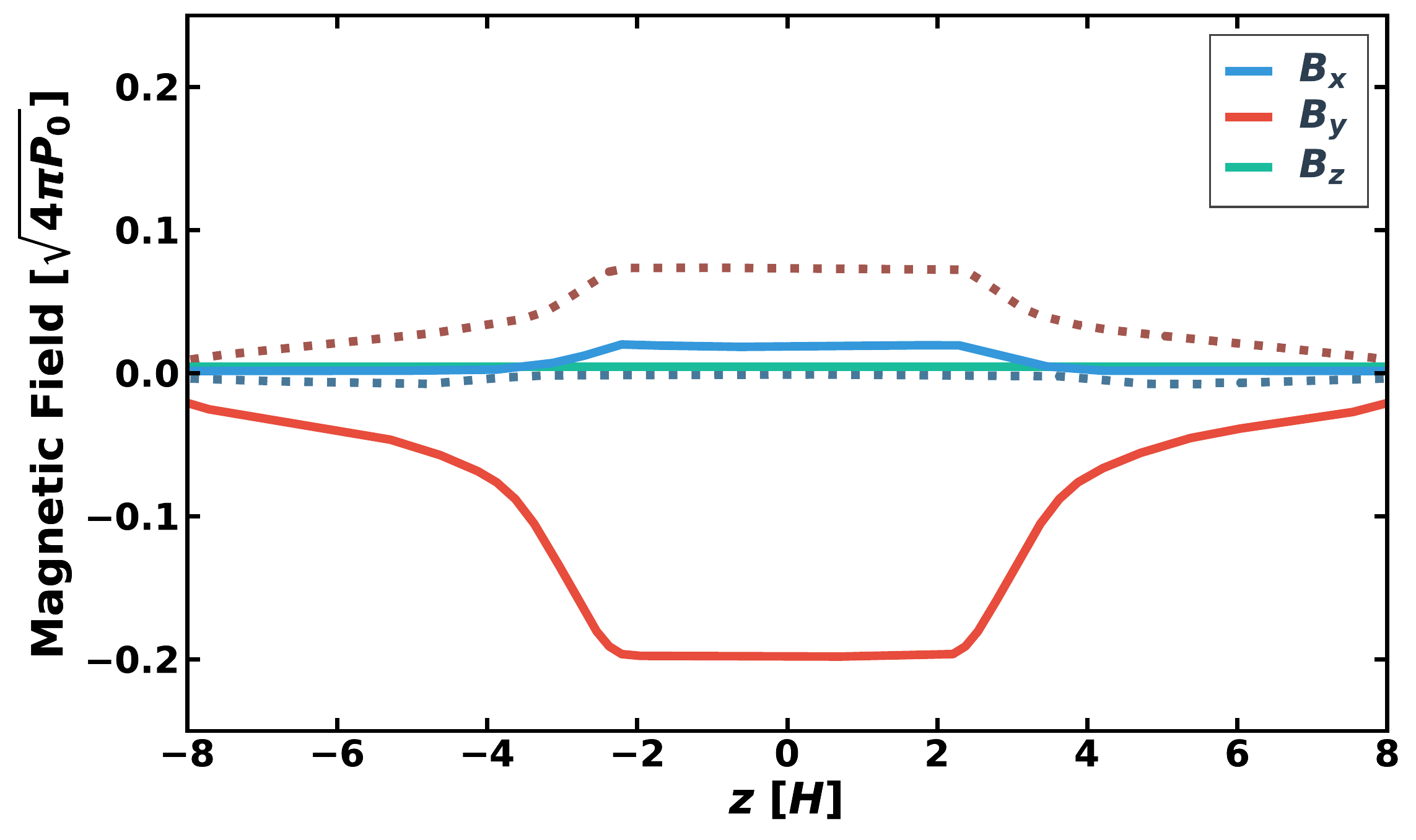}
	\includegraphics[width=0.49\hsize,clip]{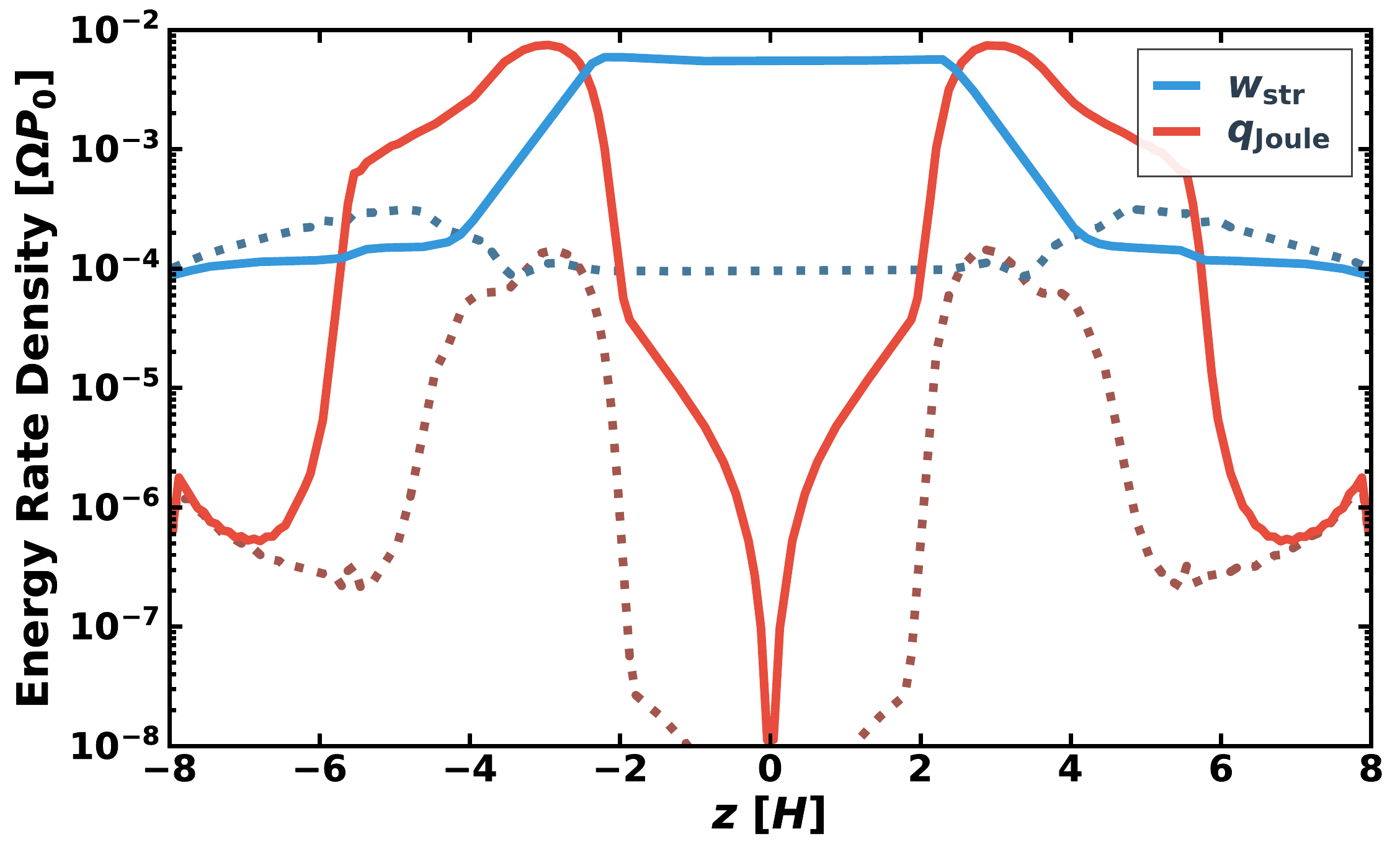} \\
	\includegraphics[width=0.49\hsize,clip]{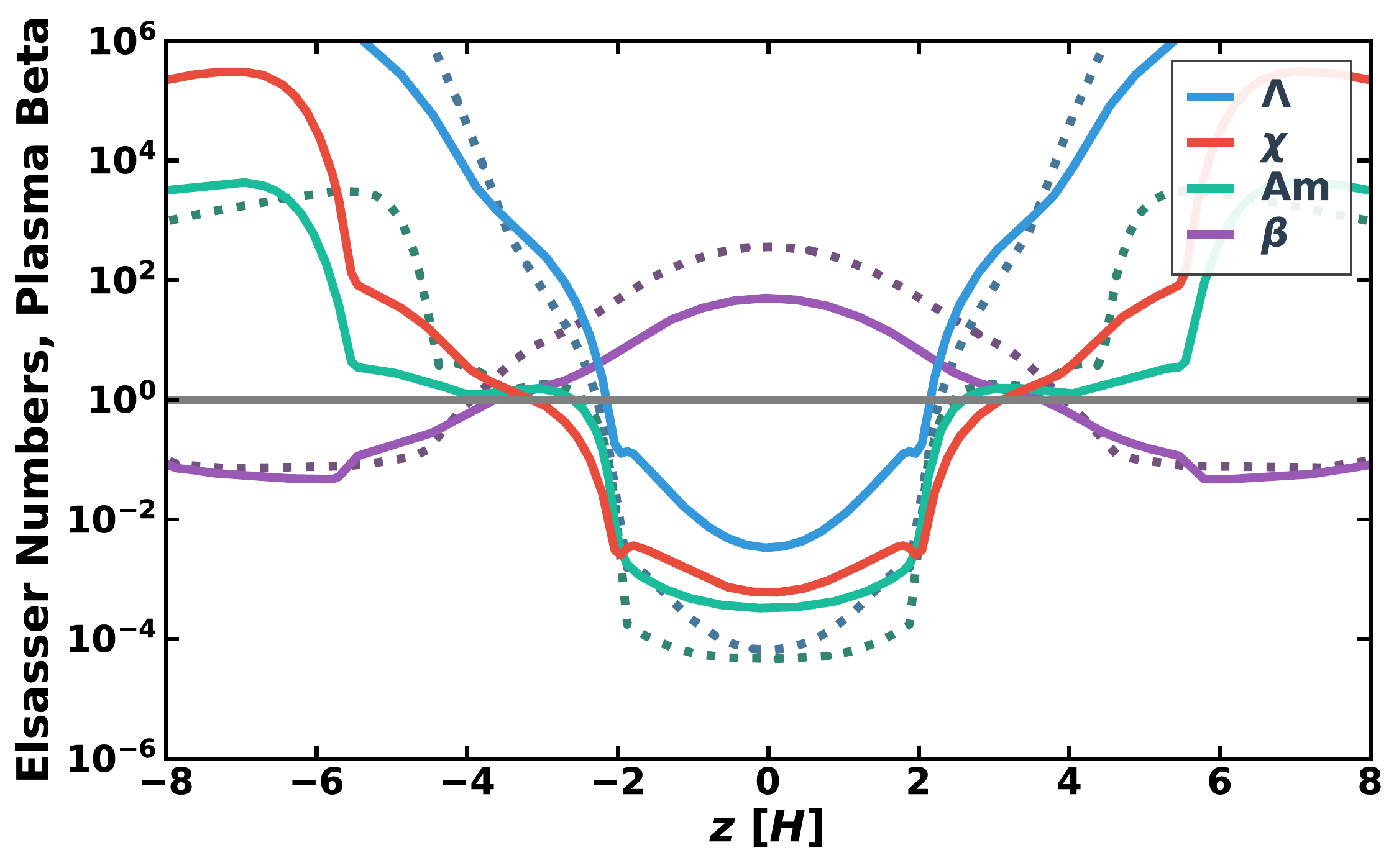}	
	\includegraphics[width=0.49\hsize,clip]{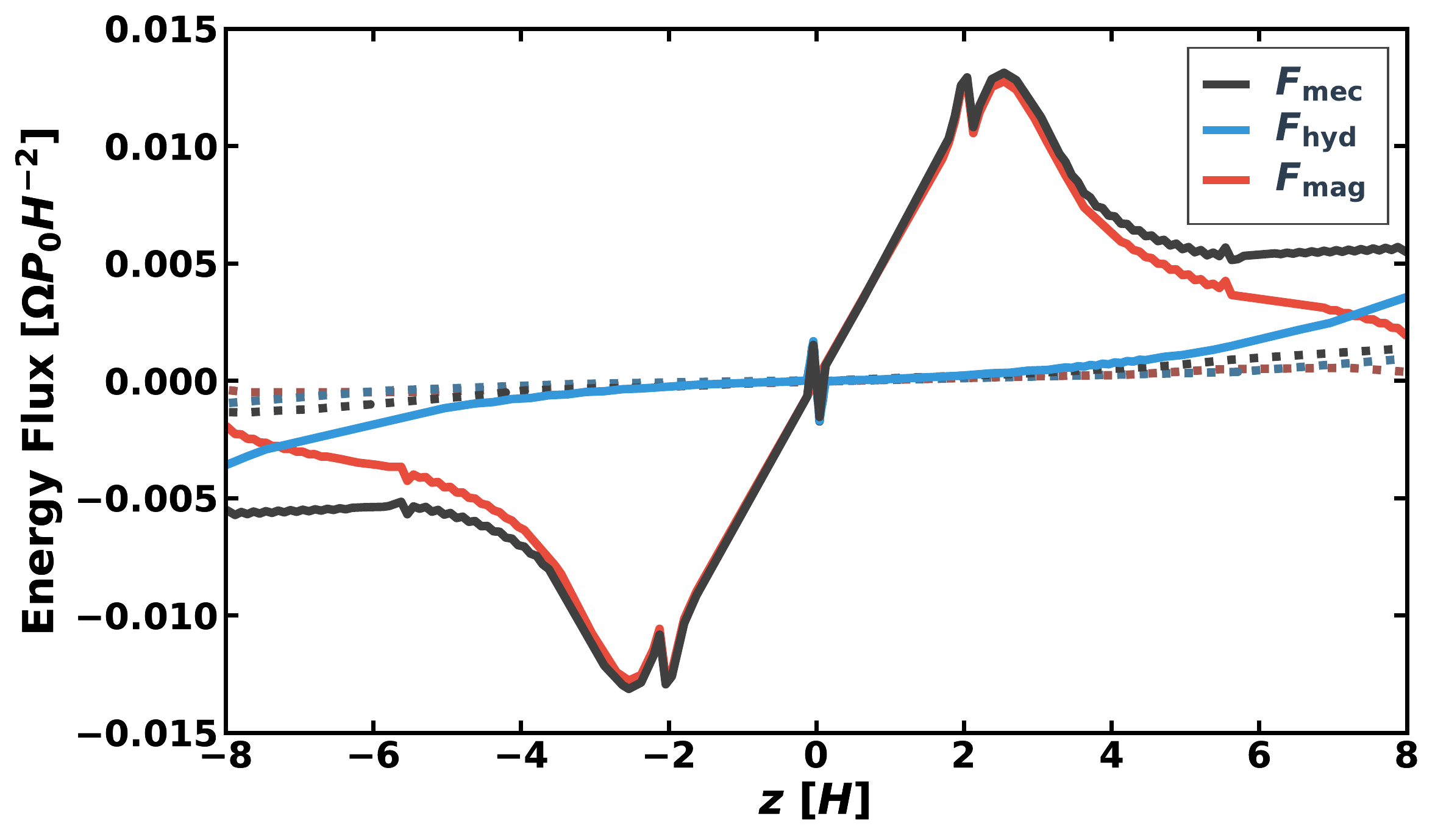}
	\caption{ 	
	Vertical profile of magnetic fields (upper left panel), 
	rates of energy injection and dissipation (upper right panel), 
	the Elsasser numbers and plasma beta (lower left panel), and the energy fluxes (lower right panel). 
	In the lower right panel, 
	$F_{\rm hyd} $ and $F_{\rm mag}$ are the hydrodynamic and magnetic energy fluxes corresponding to the first three and last three terms, respectively, in the right-hand side of \eqref{eq:mech-energy-flux}.
	The solid lines are for the fiducial simulation with all non-ideal MHD effects turned on 
and with aligned vertical field geometry $B_{z} >0$. 
	The dashed lines show results from the run without the Hall effect for comparison.}
	\label{fig:fid+}
\end{figure*}

\begin{figure*}[t]
	\centering
	\includegraphics[width=0.49\hsize,clip]{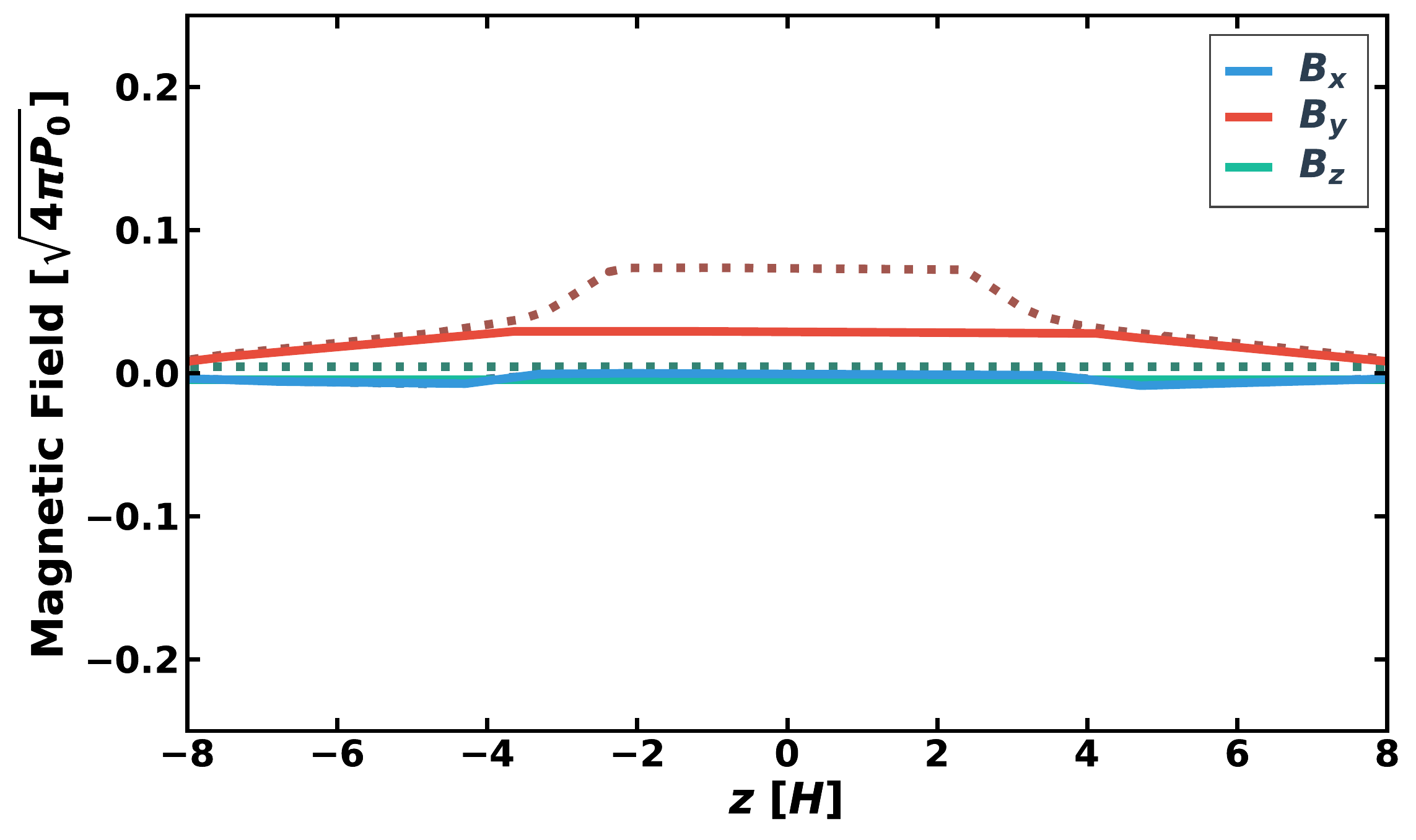}
	\includegraphics[width=0.49\hsize,clip]{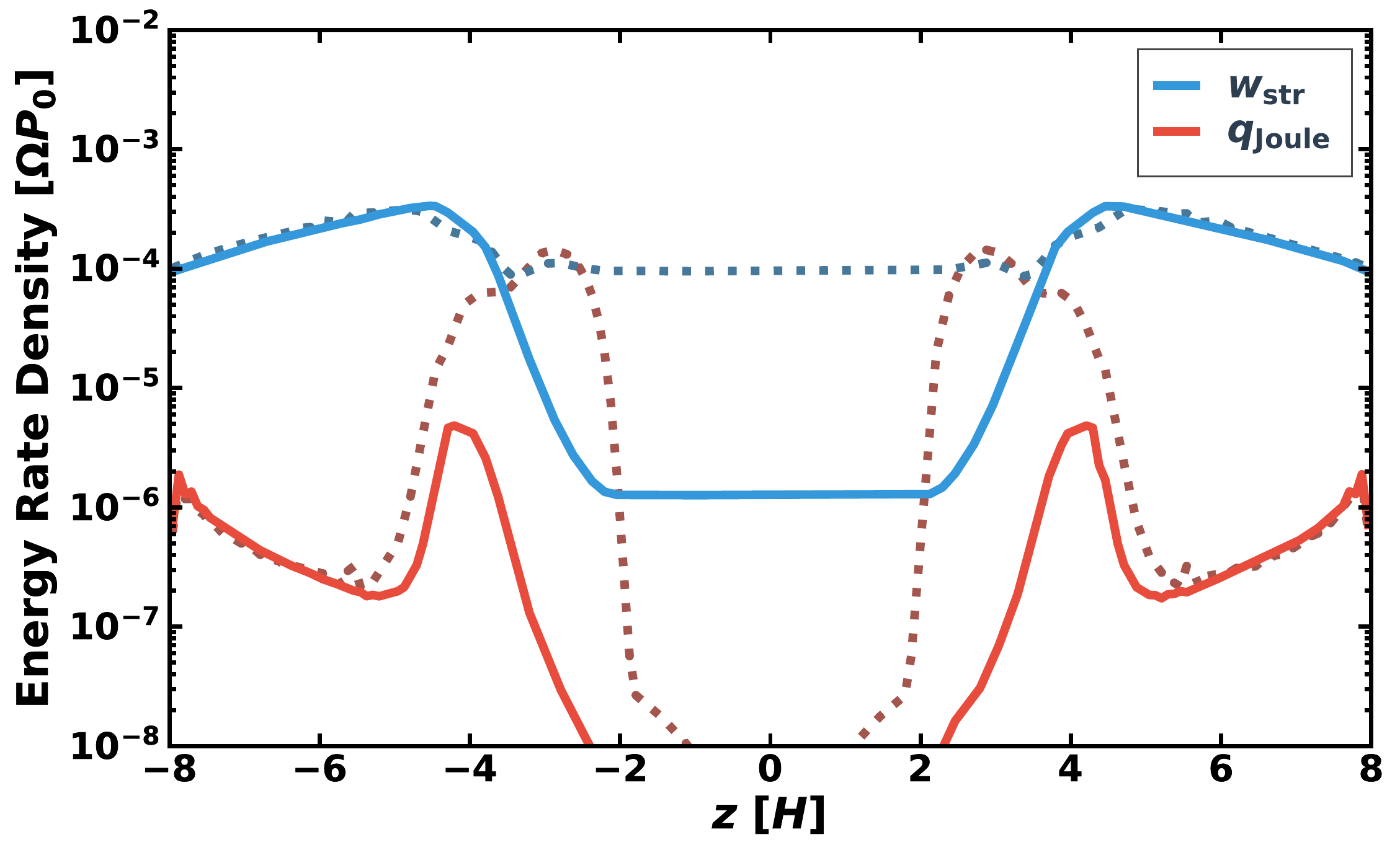}\\
	\includegraphics[width=0.49\hsize,clip]{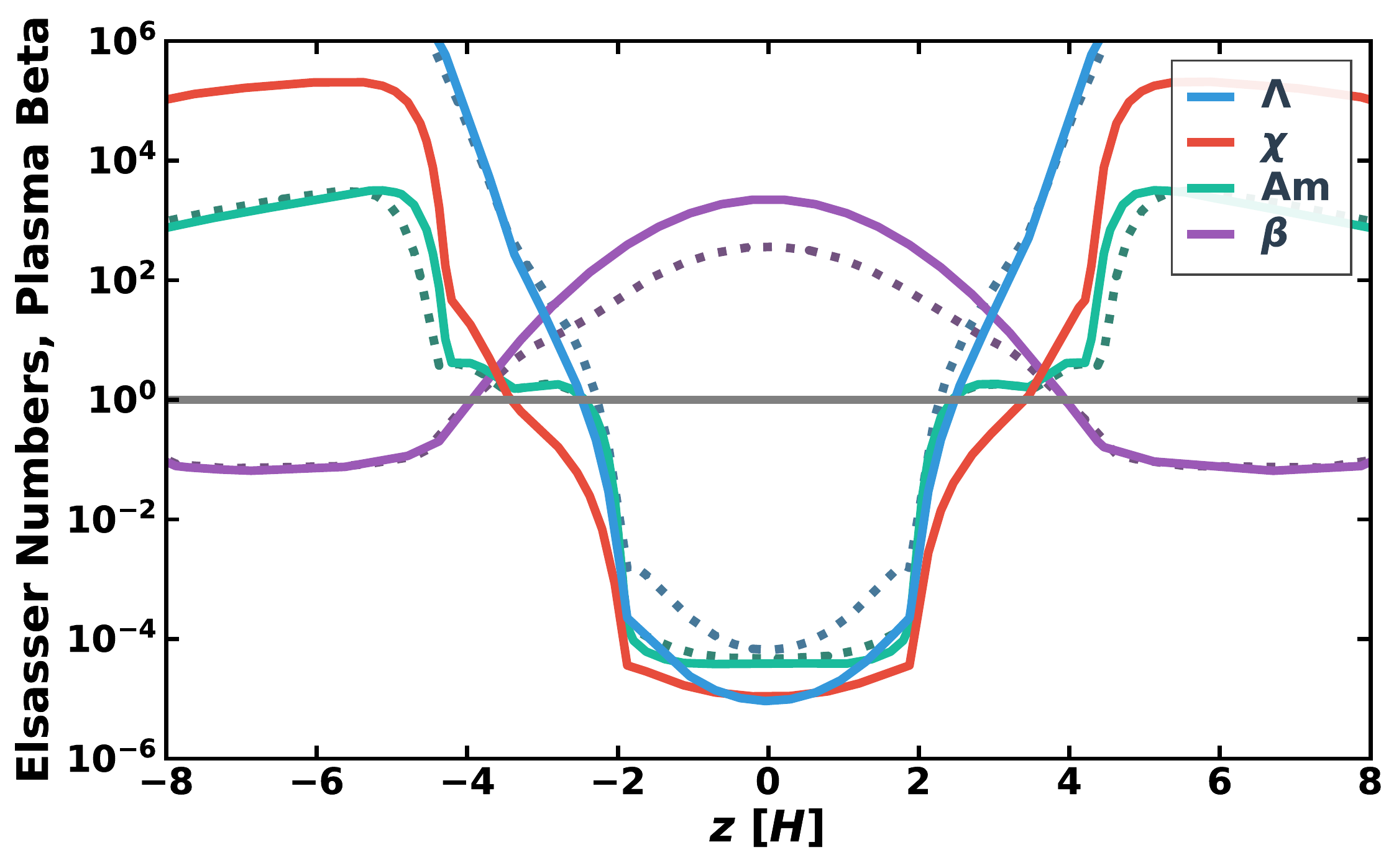}	
	\includegraphics[width=0.49\hsize,clip]{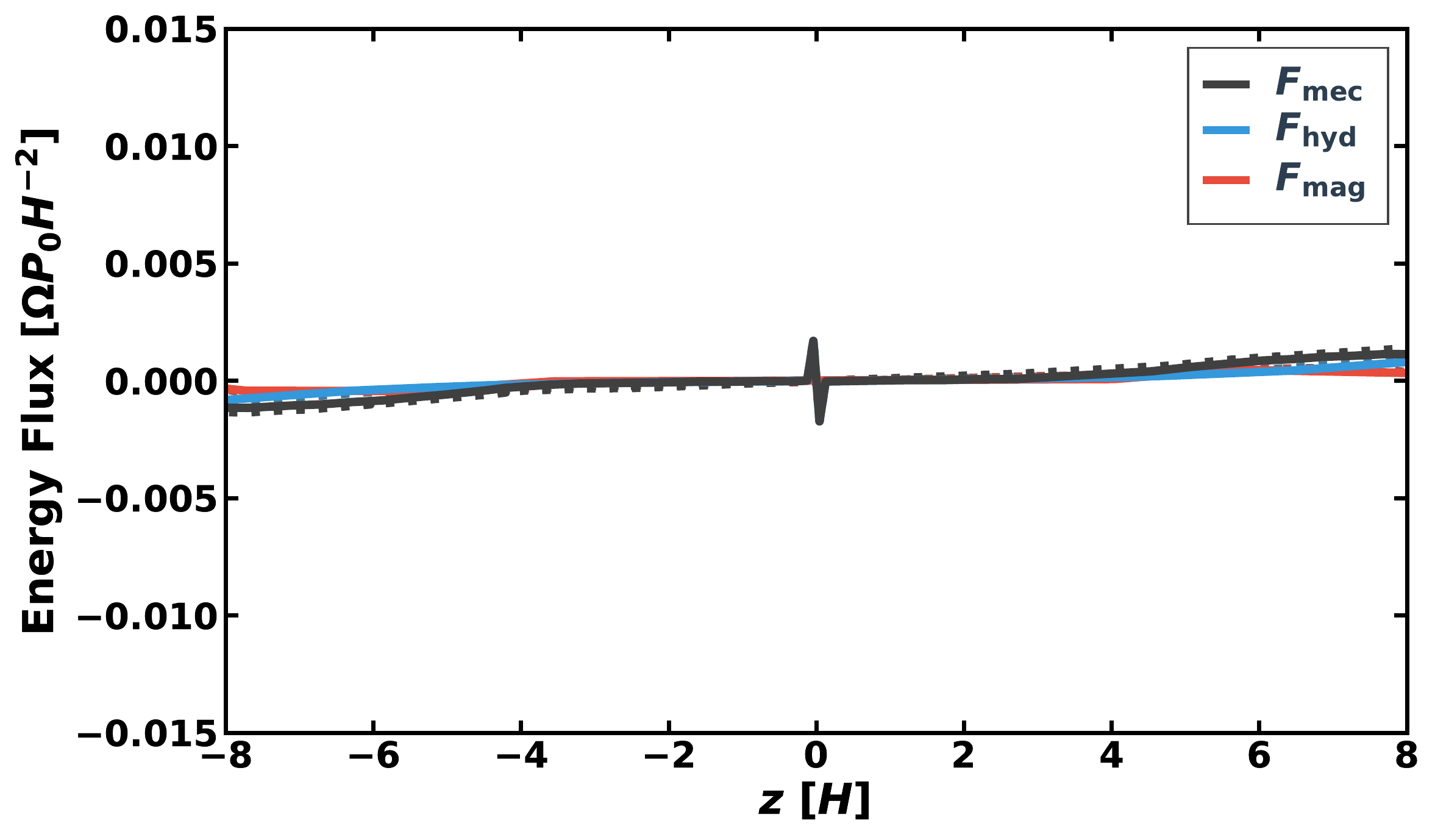}
	\caption{ 	
	Same as Figure \ref{fig:fid+}, but for the run with the anti-aligned vertical field geometry, $B_{z} < 0$.
	}
	\label{fig:fid-}
\end{figure*}

Figures \ref{fig:fid+} and \ref{fig:fid-} show the vertical profiles of the magnetic field,
the injected and dissipated energy, the Elsasser numbers, and the energy flux, for fiducial
simulations with $B_z>0$ and $B_z<0$, respectively.
The results for the same parameter set but without the Hall effect are also shown for comparison. 

The overall dynamics and magnetic field profiles are
largely controlled by non-ideal MHD effects. We see from the Elsasser number profiles in
the bottom left panel of Figure \ref{fig:fid+} and \ref{fig:fid-} that all three non-ideal
MHD effects are important within about $z=\pm2H$. 
This results from the extremely low ionization fraction, which strongly reduces the coupling
between gas and magnetic field. The lack of charge carriers also tends to yield a flat
magnetic field profile (being unable to sustain current), as seen in the corresponding
top left panels. As ionization level increases, non-ideal MHD effects weaken towards the
surface. Moreover, as density drops, ambipolar diffusion becomes the sole dominant effect,
and remains important up to $z\sim\pm4.5H$ at the location of FUV ionization front. This is
the key to MRI suppression in the disk surface \citep{Bai2013aWind-driven-Acc}. Beyond the FUV front, the
gas behaves close to the ideal MHD regime and a magnetized disk wind is launched.

\subsubsection{The Case of $B_z>0$}

We first discuss the $B_z>0$ case. On the top left panel of Figure \ref{fig:fid+}, we see
that horizontal components of the magnetic fields around the midplane region are
strongly amplified, and is also stronger than that without the Hall effect by a factor of two.
This is due to the Hall-shear instability
\citep[HSI;][]{Kunz2008On-the-linear-s,Lesur2014Thanatology-in-,Bai2014Hall-effect-Con}. This
instability simultaneously amplifies radial and toroidal fields through shear and Hall drift,
creating a strong Maxwell stress.

The profile of the magnetic field is determined by non-ideal MHD effects (amplification by HSI
within $\pm2H$ and smoothing by ambipolar diffusion) together with advection by disk winds
towards the surface. The outcome is a strong
vertical gradient of toroidal field $B_y$ beyond $\pm2H$. This gradient of $B_y$ is primarily
responsible for wind launching \citep{Bai2016Magneto-thermal}. In the mean time, it produces relatively
strong current in the disk upper layers, and leads to the energy dissipation beyond
$z\sim \pm2H$.
We see from the top right panel of Figure \ref{fig:fid+} that the energy dissipation rate
peaks at $z\sim\pm3 H$. The vertically integrated dissipation rate (Equation \ref{eq:GamJoule}), in code units, is found to be
$\Gamma_{\rm Joule}=2.4\times10^{-2}$. This is a factor of 65
higher than the
Hall-free case, which has less weaker toroidal field and its vertical gradient giving
a value of $\Gamma_{\rm Joule}=3.7\times10^{-4}$.
These values for all parameter sets are summarized in Table \ref{tab:results}.

Energy injection is dominated by the Maxwell stress, concentrated within $z\sim\pm2H$
(as a result of the HSI) as shown on the upper right panel of Figure \ref{fig:fid+}.
The vertically integrated energy injection rate (Equation \ref{eq:Winj}) reaches $W_{\rm str}=3.3\times10^{-2}$,
much higher than the Hall-free case, which gives $W_{\rm str}=2.5\times10^{-3}$.
From the bottom right panel of the figure, we further see that this energy is carried to
upper layers first by the Poynting flux of ambipolar diffusion, $\etaA \bm{J} \times \bm{B}$, and then by advection
through disk wind.\footnote{There are a few spikes in the energy flux profile.
The spikes at the midplane are caused by a numerical error of the gravitational potential,
and the ones at $z\sim\pm2 H$ are caused by switching on the diffusivity cap.
These spikes are in very limited regions and do not affect the overall results.}

The overall conservation of mechanical energy is achieved in the simulations. As
discuss in Appendix \ref{sec:cons}, part of the energy injection by Maxwell stress
is dissipated into $q_{\rm Joule}$, while the rest is used to drive disk winds.
The division of $w_{\rm str}$ into the two parts depends on disk microphysics.
Although the Hall effect does not generate the Joule heating, the HSI amplifies the
magnetic field and hence enhances the Joule heating. The conversion of the work done
by Maxwell stress into Joule heating reaches $\sim71\%$ for this fiducial run,
as opposed to only $\sim 15\%$ in the Hall-free run.

Finally, we comment that in this simulation, the toroidal magnetic field takes the same
sign over the entire computational domain. This geometry is, however, unphysical for
wind launching in a global disk \citep{Bai2013aWind-driven-Acc}. This is a main limitation
of local simulations where there is no preference of being radially inward or outward.
The influence of the unphysical field geometry on the temperature profile is discussed in Section \ref{ssec:disc-sym}.

\subsubsection{The Case of $B_{z}<0$}\label{sec:bz-}

While the Hall effect amplifies horizontal magnetic field when $B_{z}>0$, it suppresses the growth of the horizontal field when $B_{z}<0$ \citep[e.g.,][]{Bai2014Hall-effect-Con}.
Figure \ref{fig:fid-} shows the same as Figure \ref{fig:fid+}, but for run with $B_{z} < 0$.
We see that horizontal magnetic field strength is maintained at a relatively low level within $\pm 4 H$,
as compared to the Hall-free case.
Consequently, the current generated by the vertical gradient of $B_{y}$ is weaker, leading to much
smaller Joule heating rate even compared with the Hall-free case.
As a result, Joule heating is much weaker, and we find $\Gamma_{\rm Joule}$ to be only $9.5\times10^{-6}$.

The work done by the Maxwell stress is also significantly lower around the midplane
region within $z=\pm2H$ (again consequence of the Hall effect suppressing horizontal
field). Towards disk upper layers, as the Hall effect significantly weakens, we see
that the magnetic field profiles are almost identical to the Hall-free case, which
still yields considerable Maxwell stress, amounting to $W_{\rm str}=1.9\times10^{-3}$,
which is associated with the wind launching process. In other words, most of the
Maxwell stress is generated to assist wind launching instead of Joule dissipation. 

\subsection{Temperature Profiles}\label{ssec:res-temp}

\begin{figure*}[t]
	\centering
	\includegraphics[width=0.49\hsize,clip]{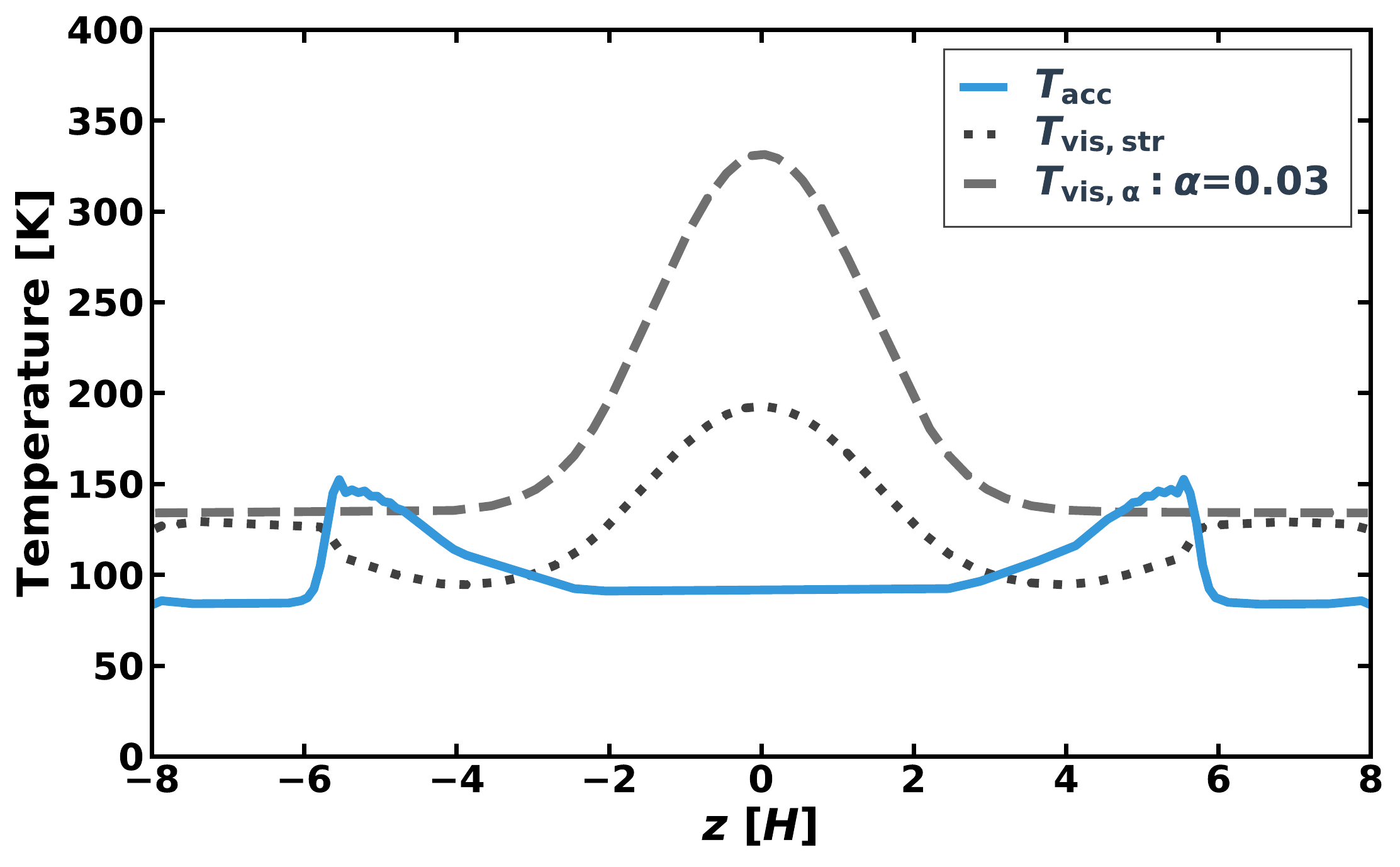}
	\includegraphics[width=0.47\hsize,clip]{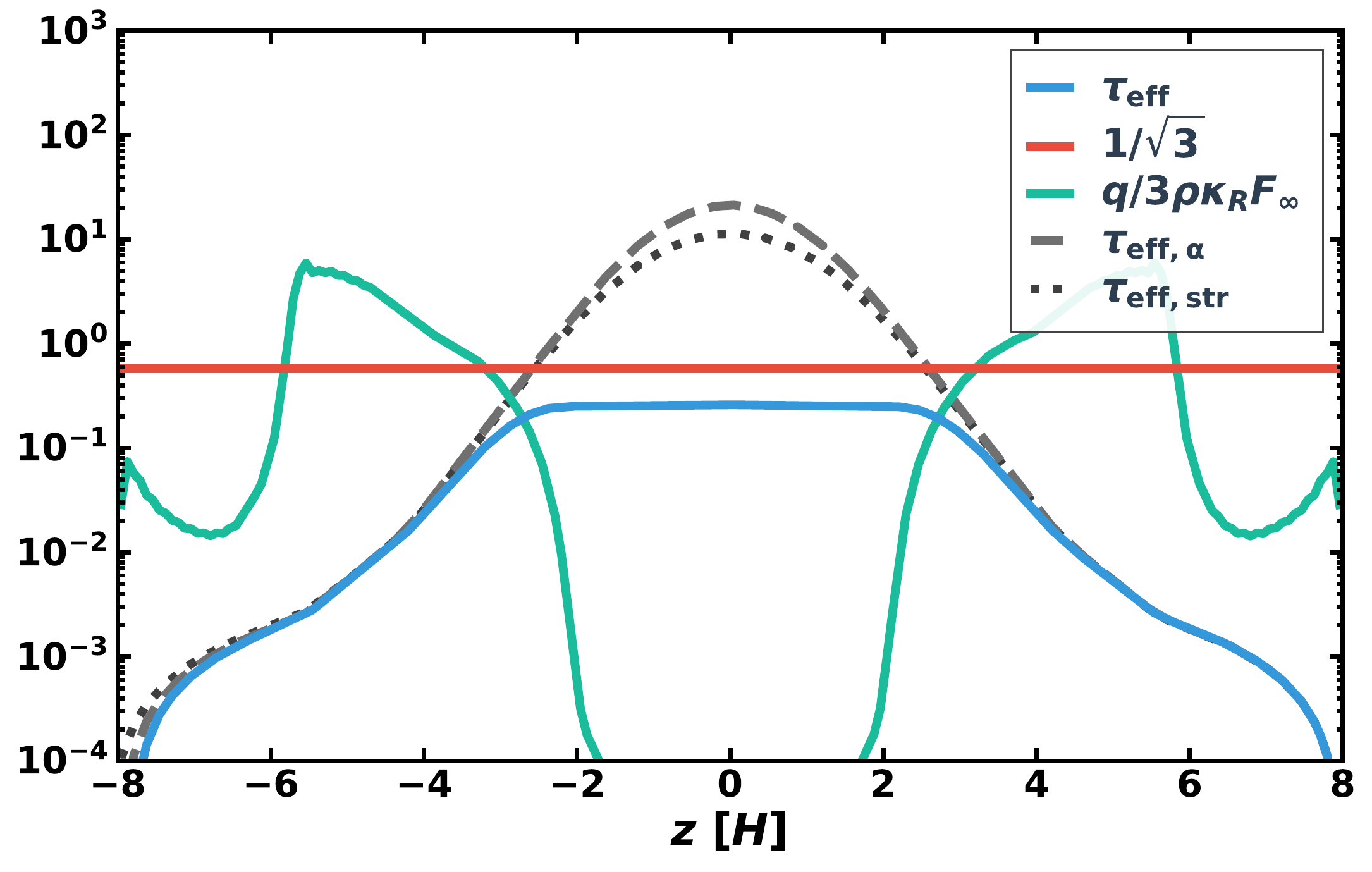}\\
	\includegraphics[width=0.49\hsize,clip]{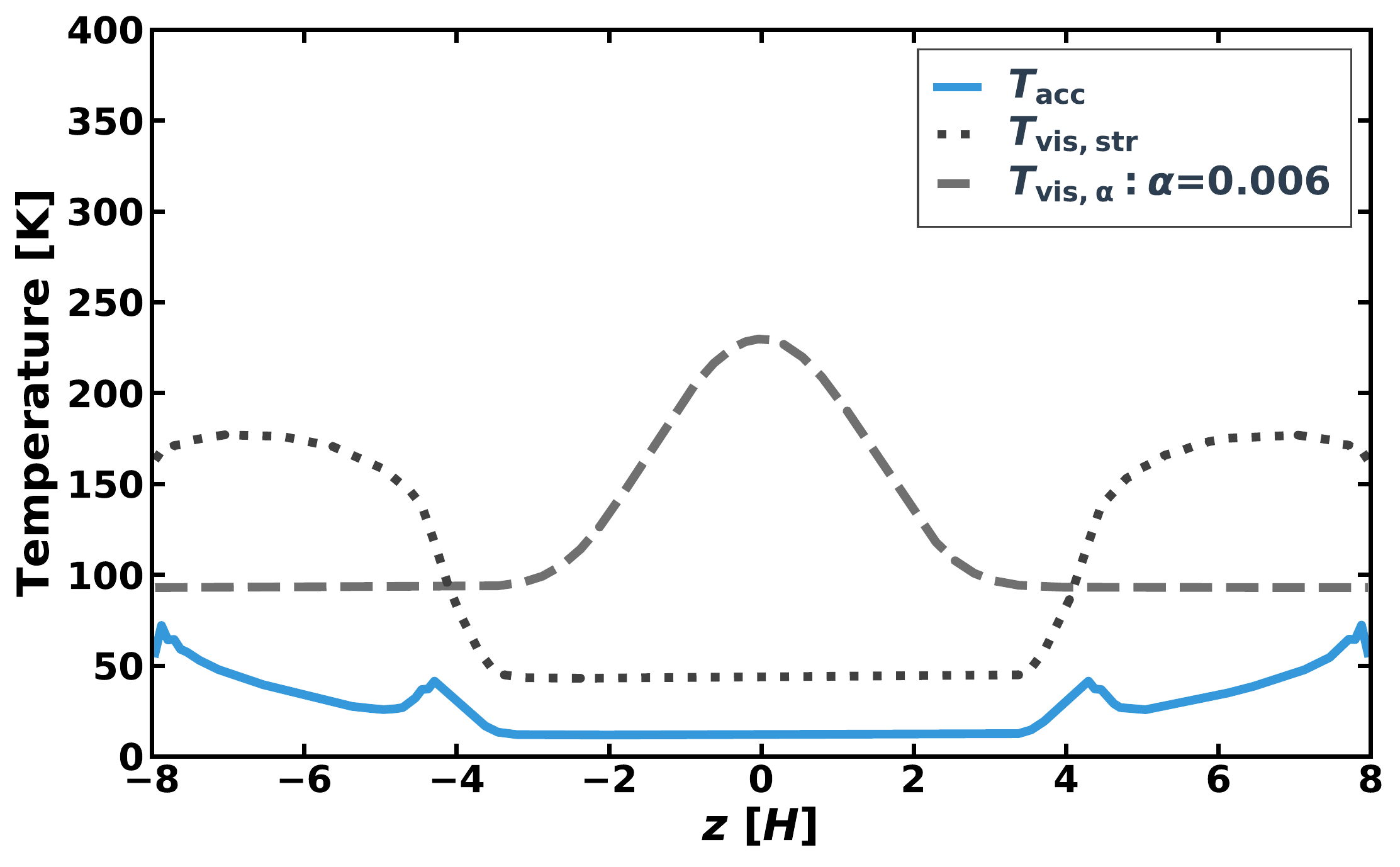}
	\includegraphics[width=0.47\hsize,clip]{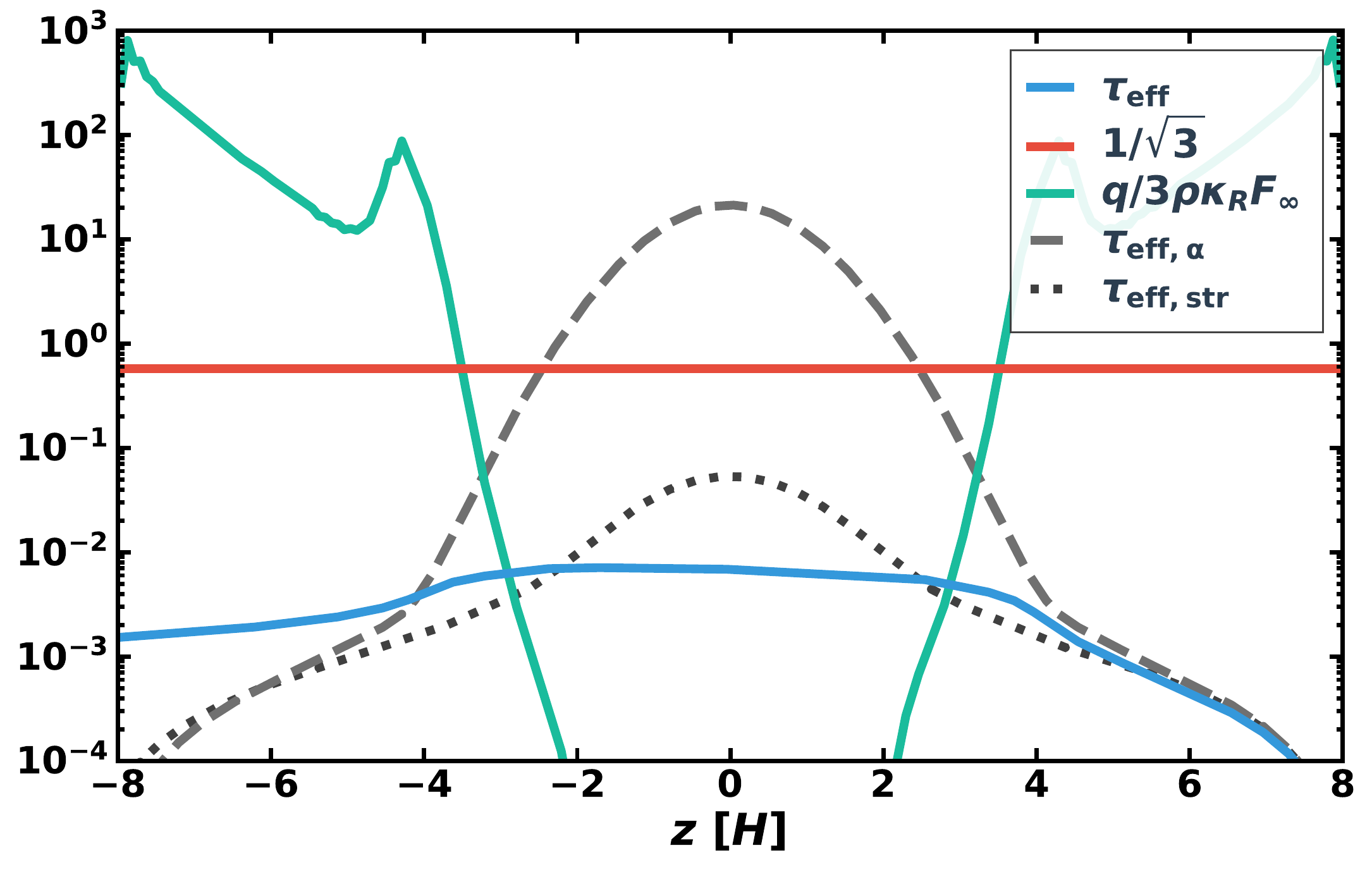}
	\caption{  
	Vertical profiles of temperature (left panels) and effective optical depth (right panels) for the runs with $B_z>0$ (upper panels) and $B_z<0$ (lower panels).
	The solid line in the left panels shows the temperature profiles derived by using the Joule dissipation rate from the MHD simulations.
	The dotted and dashed lines show the temperature profiles derived from the ``equivalent" viscous model, $T_{\rm vis,\,str}$, and the `constant-$\alpha$" model $T_{\rm vis,\,\alpha}$ (see Section \ref{ssec:temp}), respectively.
	Irradiation heating is not included.
	In the right panels, the blue lines are $\tau_{\rm eff}$, while the red and green lines are 
	the second and third terms in the parentheses of \eqref{eq:T}, respectively, which helps compare their relative importance in determining $T(z)$. 
	The dotted and dashed lines in the right panels show the effective optical depth for the equivalent viscous model, $\tau_{\rm eff,\,str}$, and constant-$\alpha$ model, $\tau_{\rm eff,\,\alpha}$, respectively. 
	}
	\label{fig:temp-vs-old}
\end{figure*}

In this subsection, we reconstruct the vertical temperature profile based on the heating rate obtained earlier.

\subsubsection[]{No irradiation}

First, we focus on temperature profiles determined only by the accretion heating. Without considering irradiation, the typical temperatures found in the calculations are smaller or even much smaller than the temperature assumed in our simulations. Here, we mainly focus on the comparison between different heating prescriptions (accretion heating by Joule dissipation, equivalent viscous dissipation, and constant $\alpha$), and different field geometries. 

Figure \ref{fig:temp-vs-old} shows that the temperature from the simulation and two viscous heating models for runs with the $B_z>0$ and $B_z<0$ cases.
We also show the profile of the effective optical depth (see \eqref{eq:taueff} and right panels of Figure \ref{fig:temp-vs-old}),
which measures the optical depth above a certain height weighted by the heating profile. 
The effective optical depth encapsulates the crucial differences among heating profiles.
With Joule heating, and for both $B_z>0$ and $B_z<0$ cases, since the heating occurs at $z\approx\pm3H$ and $\pm4H$, respectively, the effective optical depth no longer increases within a few scale heights. This yields relatively low temperature at the midplane region, and temperature peaks towards disk surface where most heating takes place.
Overall, the temperature in the $B_z>0$ case is much larger than that for $B_z<0$, which is largely due to the different level of total Joule dissipation controlled by the Hall effect.

In the equivalent viscous model, there is strong ($B_z>0$ case) and modest ($B_z<0$ case) energy release in the midplane region, making the effective optical depth peaking in the midplane, together with higher midplane temperatures, especially in the $B_z>0$ case. Temperature increases further in the surface again due to the strong Maxwell stress there.

In the constant-$\alpha$ model, the total heating rate is much higher than other models, with heating profile centrally peaked. This leads to a centrally peaked temperature profile with significantly higher midplane temperature. Note that in the simulations, wind-driven accretion dominates in both $B_z>0$ and $B_z<0$ cases with similar total accretion rates. Therefore, in the constant-$\alpha$ model (with $\alpha$ value chosen so that the resulting viscous accretion rate matches that from the simulations), both the effective optical depth and the resulting temperature profiles are similar in the two cases. 
Interestingly, despite much stronger total heating rate, the constant-$\alpha$ model generally gives surface temperatures lower than the midplane temperature, 
because local heating at the surface in this model diminishes as density drops.

As constant-$\alpha$ models have been widely used in the literature, the dramatic difference between the resulting temperature profiles and those obtained from our simulation results demonstrate the importance of better understanding the energy dissipation in disks.
Hereafter, comparisons will be made only with the constant-$\alpha$ model.

\begin{figure*}[t]
	\centering
	\includegraphics[width=\hsize,clip]{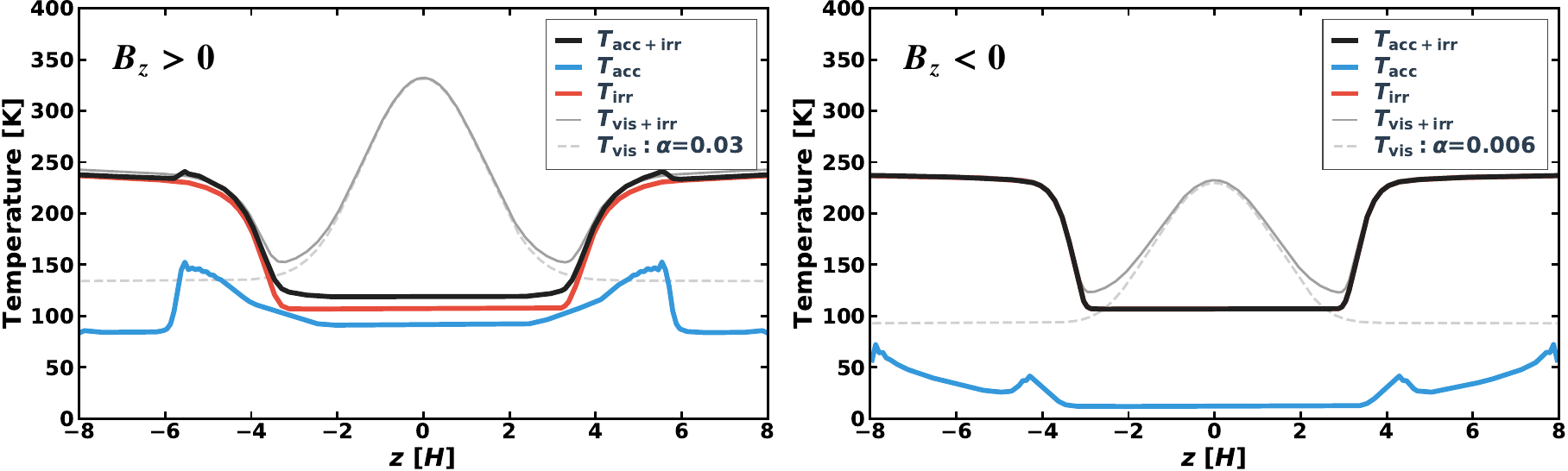}
	\caption{
	Vertical temperature profiles computed taking into account accretion heating only (blue lines), irradiation heating only (red lines), and both contributions (black lines)
	for the cases of $B_{z}>0$ (left panel) and $B_{z}<0$ (right panel). 
	For comparison, the temperature profiles from the constant-$\alpha$ model with and without irradiation heating are also shown as gray solid and dashed lines, respectively.}
	\label{fig:temp}
\end{figure*}

\subsubsection[]{With irradiation}

We add heating energy rate $q_{\rm irr}$ of the stellar irradiation flux into
$q=q_{\rm Joule}+q_{\rm irr}$, and then calculate the temperature profile determined by both
the irradiation and Joule heating in the same way as before.
In doing so, we assume that the stellar irradiation and disk thermal radiation are well-separated
radiation fields in wavelength, allowing us to solve them separately \citep{Calvet1991Irradiation-of-,Guillot2010On-the-radiativ}. 
The stellar irradiation flux is given by \citep{Calvet1991Irradiation-of-}
\begin{equation}
	\mathcal{F}_{\rm irr} (z) = - E_{0}\mu_{0} \pr{ \exp\pr{ - \frac{ \tau_{\rm vi}(z) }{  \mu_{0} } } +   \exp\pr{ - \frac{ \tau_{\rm vi}(-\infty) -\tau_{\rm vi}(z)  }{  \mu_{0} } }   } \ ,
\end{equation}
where 
\begin{eqnarray}
	\mu_{0} = r \frac{\d}{\dr}\pr{\frac{H_{\rm p}}{r}}
\end{eqnarray}
is cosine of the angle from stellar incident flux to normal of the disk surface,
$R_{*}$ is the stellar radius,
$H_{\rm p}$ is the height of the photosphere, 
$E_{0}$ is the incoming energy flux at the disk surface, and
\begin{equation}
	\tau_{\rm vi}(z) = \int_{z}^{+\infty} \rho \kappa_{\rm vi} \dz
\end{equation}
is the optical depth for visible light, where $\kappa_{\rm vi}$ is the opacity for visible light.
In this paper, we assume $\kappa_{\rm vi}=\kappa_{\rm R}$.
We take $R_{*}$ to be the solar radius, and $H_{\rm p}$ to be $4H$.
We also take $E_{0}$ to be $L_{\sun}/(8 \pi r^{2})$, where we assume that stellar irradiation comes from one side of the star and $L_{\sun}$ is the solar luminosity.
The rate profile of heating energy of the stellar irradiation flux is then given by
\begin{eqnarray}
q_{\rm irr} &=& - \pd{\mathcal{F}_{\rm irr}}{z}  \nonumber \\
	 &=& E_{0} \rho \kappa_{\rm vi} \pr{ \exp\pr{ - \frac{ \tau_{\rm vi}(z) }{  \mu_{0} } } +   \exp\pr{ - \frac{ \tau_{\rm vi}(-\infty) -\tau_{\rm vi}(z)  }{  \mu_{0} } }   } \ . 
\end{eqnarray}

Figure \ref{fig:temp} shows the temperature profiles taking into account the accretion heating (Joule heating), the irradiation heating, and the both.
We find that for both $B_{z}>0$ and $B_{z}<0$ cases, 
the temperature profile is primarily determined by irradiation. Contribution from Joule heating is much smaller. With $B_{z}>0$ where Joule dissipation is stronger, disk midplane temperature is only enhanced by a small fraction, with additional small temperature enhancement at the surface up to about 5 scale heights where Joule dissipation profile peaks.
In the case of $B_{z}<0$, the Joule heating is so weak that its contribution to the temperature profile
is largely negligible. 
This is in strong contrast with the constant-$\alpha$ model, where the disk midplane temperature is fully dominated by viscous heating, making midplane temperature much higher.

\newcommand{\mpyr}{ [M$_{\sun}$/yr]}
\newcommand{\Ke}{ [K]}

\begin{deluxetable*}{rrrr|rrrrrrrrrr}
\tabletypesize{\footnotesize}
\tablecaption{Summary of the results for all parameter sets \label{tab:results}}
\tablehead{ \colhead{$r$ [AU]} & \colhead{$\Sigma$ [g/cm$^{2}$]}  & \colhead{$\beta_{0}$} & \colhead{$f_{\rm dg}$} &
		  \colhead{$\alpha_{r}$}& \colhead{$\alpha_{z}$} & 
		  \colhead{$\dot{M}_{r}$\mpyr}& \colhead{$\dot{M}_{z}$\mpyr}&
		  \colhead{$W_{\rm str}$}& \colhead{$\Gamma_{\rm Joule}$}&	\colhead{$\Gamma_{\rm acc}$} &	  
		   \colhead{$T_{\rm acc}$\Ke} & \colhead{$T_{\rm vis}$\Ke} & \colhead{$T_{\rm irr}$\Ke}}
\startdata
1 & 1700 & 1e5 & 1e-4   & 1.1e-2 & 1.6e-2 & 5.7e-8 & 8.0e-8    & 3.3e-2 & 2.4e-2 & 1.5e-1   &  91 & 330 & 105 \\
1 & 1700 & 1e5$^{*}$ & 1e-4   & 2.9e-4 & 7.3e-3 & 1.4e-9 & 3.6e-8    & 2.5e-3 & 3.7e-4 & 4.3e-2   &  32 & 239 & 105 \\
1 & 1700 & -1e5 & 1e-4   & 5.5e-5 & 6.4e-3 & 2.7e-10 & 3.2e-8    & 1.9e-3 & 9.5e-6 & 3.6e-2   &  12 & 229 & 105 \\
\hline
1 & 17000 & 1e5 & 1e-4   & 4.5e-3 & 1.2e-2 & 2.3e-7 & 5.9e-7    & 1.4e-2 & 9.3e-3 & 9.3e-2   & 135 & 911 & 106 \\
1 & 17000 & -1e5 & 1e-4   & 1.2e-6 & 2.6e-3 & 6.1e-11 & 1.3e-7    & 5.5e-4 & 3.8e-5 & 1.5e-2   &  32 & 573 & 105 \\
1 & 170 & 1e5 & 1e-4   & 1.8e-2 & 2.4e-2 & 9.2e-9 & 1.2e-8    & 5.4e-2 & 3.6e-2 & 2.4e-1   &  56 & 124 & 105 \\
1 & 170 & -1e5 & 1e-4   & 7.7e-4 & 1.2e-2 & 3.8e-10 & 6.1e-9    & 7.0e-3 & 7.2e-4 & 7.3e-2   &  20 &  92 & 105 \\
1 & 17 & 1e5 & 1e-4   & 1.0e-2 & 5.4e-2 & 5.0e-10 & 2.7e-9    & 3.9e-2 & 9.9e-3 & 3.6e-1   &  23 &  67 & 106 \\
1 & 17 & -1e5 & 1e-4   & 6.3e-3 & 3.8e-3 & 3.2e-10 & 1.9e-10    & 6.0e-2 & 7.4e-3 & 5.7e-2   &  20 &  42 & 106 \\
\hline
2.0 & 601 & 1e5 & 1e-4   & 1.2e-2 & 1.4e-2 & 4.5e-8 & 5.1e-8    & 3.5e-2 & 2.4e-2 & 1.5e-1   &  51 & 140 &  78 \\
2.0 & 601 & -1e5 & 1e-4   & 9.1e-4 & 2.5e-2 & 3.4e-9 & 9.3e-8    & 1.1e-2 & 4.2e-3 & 1.5e-1   &  33 & 140 &  77 \\
0.5 & 4808 & 1e5 & 1e-4   & 1.0e-2 & 2.1e-2 & 6.7e-8 & 1.4e-7    & 3.0e-2 & 2.1e-2 & 1.7e-1   & 162 & 793 & 146 \\
0.5 & 4808 & -1e5 & 1e-4   & 3.8e-5 & 5.8e-3 & 2.5e-10 & 3.9e-8    & 1.3e-3 & 2.5e-5 & 3.3e-2   &  27 & 524 & 146 \\
0.2 & 19007 & 1e5 & 1e-4   & 7.4e-3 & 2.9e-2 & 7.3e-8 & 2.9e-7    & 2.3e-2 & 1.5e-2 & 2.0e-1   & 340 & 2549 & 235 \\
0.2 & 19007 & -1e5 & 1e-4   & 1.6e-6 & 4.2e-3 & 1.6e-11 & 4.2e-8    & 5.9e-4 & 5.2e-5 & 2.4e-2   &  77 & 1492 & 235 \\
\hline
1 & 1700 & 1e6 & 1e-4   & 3.6e-3 & 3.6e-3 & 1.8e-8 & 1.8e-8    & 1.0e-2 & 7.5e-3 & 4.1e-2   &  68 & 236 & 105 \\
1 & 1700 & -1e6 & 1e-4   & 5.8e-5 & 9.0e-4 & 2.9e-10 & 4.5e-9    & 9.6e-4 & 3.2e-5 & 5.4e-3   &  17 & 142 & 105 \\
1 & 1700 & 1e4 & 1e-4   & 3.3e-2 & 7.6e-2 & 1.6e-7 & 3.8e-7    & 1.0e-1 & 6.6e-2 & 6.1e-1   & 118 & 465 & 106 \\
1 & 1700 & -1e4 & 1e-4   & 3.3e-4 & 2.2e-2 & 1.6e-9 & 1.1e-7    & 6.3e-3 & 3.9e-4 & 1.3e-1   &  30 & 313 & 105 \\
1 & 1700 & 1e3 & 1e-4   & 8.8e-2 & 3.8e-1 & 4.4e-7 & 1.9e-6    & 3.0e-1 & 1.8e-1 & 2.6   & 152 & 670 & 106 \\
1 & 1700 & -1e3 & 1e-4   & 5.8e-3 & 9.0e-2 & 2.9e-8 & 4.5e-7    & 3.3e-2 & 1.7e-2 & 5.4e-1   &  77 & 451 & 105 \\
\hline
1 & 1700 & 1e5 & 1e-3   & 4.1e-3 & 1.3e-2 & 2.0e-8 & 6.2e-8    & 1.3e-2 & 7.2e-3 & 9.4e-2   &  81 & 514 & 106 \\
1 & 1700 & -1e5 & 1e-3   & 5.4e-5 & 6.4e-3 & 2.7e-10 & 3.2e-8    & 1.9e-3 & 9.5e-6 & 3.6e-2   &  12 & 405 & 105 \\
1 & 1700 & 1e5 & 1e-5   & 1.9e-2 & 1.8e-2 & 9.4e-8 & 9.2e-8    & 5.4e-2 & 4.1e-2 & 2.1e-1   &  97 & 213 & 105 \\
1 & 1700 & -1e5 & 1e-5   & 5.3e-5 & 6.4e-3 & 2.6e-10 & 3.2e-8    & 1.9e-3 & 9.4e-6 & 3.6e-2   &  12 & 138 & 105 \\
\enddata
\tablenotetext{*}{Run without Hall effect.}
\tablecomments{
The sign of $\beta_{0}$ express the sign of $B_{z}$.
The accretion rates $\dot{M}_{r}$ and $\dot{M}_{z}$ are calculated by the first and second terms of \eqref{eq:maccr}, respectively.
The alpha value $\alpha_r$ and $\alpha_z$ are the equivalent viscous $\alpha$ values to yield accretion rates of $\dot{M}_{r}$ and $\dot{M}_{z}$, respectively.
The energy production rate $W_{\rm str}$, $\Gamma_{\rm Joule}$, and $\Gamma_{\rm acc}$ are given by the integration of the work done by the stress $w_{\rm str}$, Joule dissipation rate $q_{\rm Joule}$, and the energy dissipation rate in the constant-$\alpha$ model to yield total accretion rate $\dot{M}_{r} + \dot{M}_{z}$. 
The temperatures $T_{\rm acc}$, $T_{\rm vis}$, and $T_{\rm irr}$ are the midplane temperatures given by the Joule heating, viscous heating of the expected mass accretion rate (from the constant-$\alpha$ model), and irradiation heating, respectively.} 
\end{deluxetable*}

\section{Parameter Exploration}\label{sec:ps}

\begin{figure*}[t]
	\centering
	\includegraphics[width=\hsize,clip]{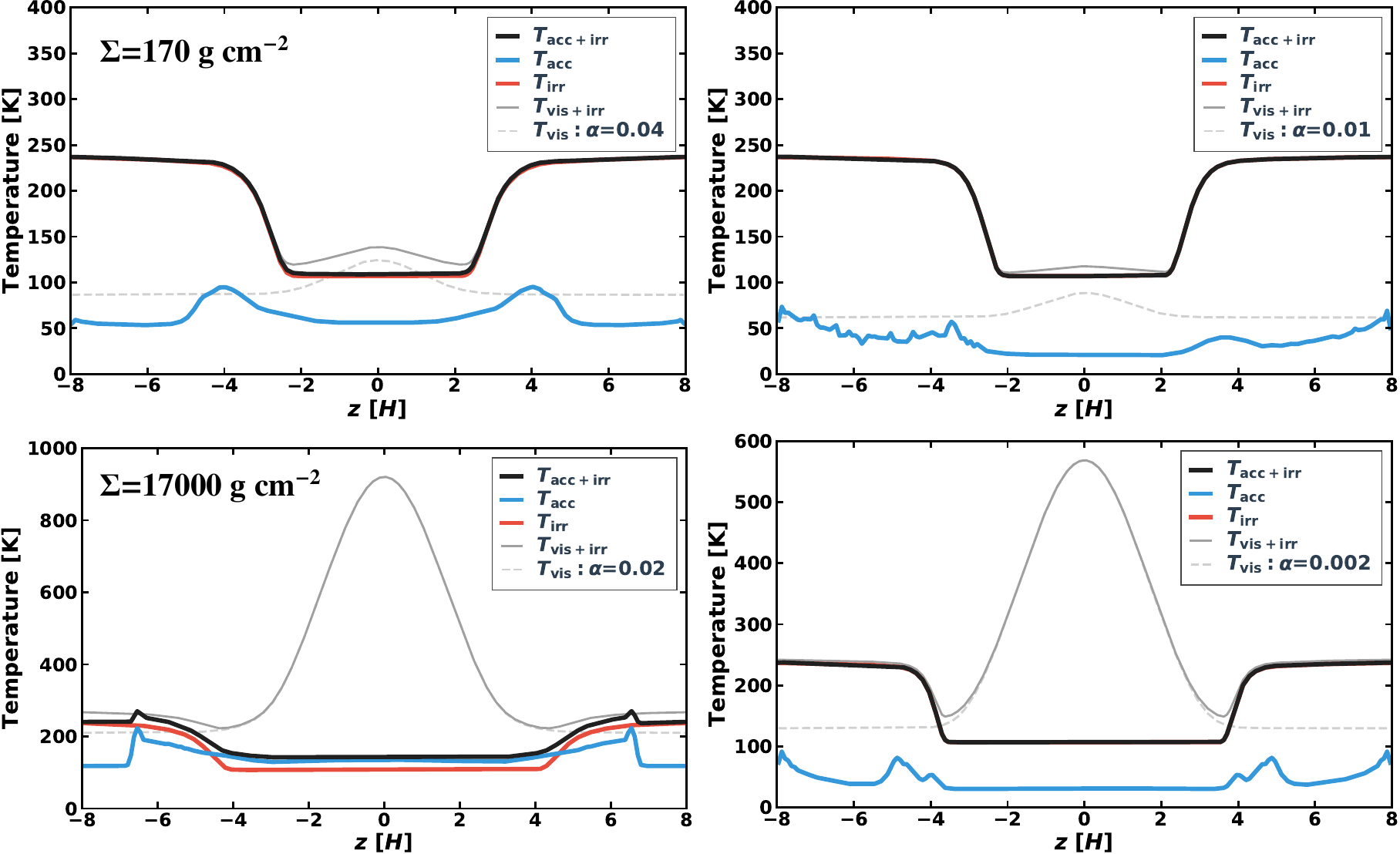}
	\caption{  	
	Same as Figure \ref{fig:temp}, but for runs with $\Sigma=170$ (top) and $17000$ g cm$^{-2}$ (bottom) and with $B_{z}>0$ (left) and $B_{z}<0$ (right). Note that the scales in some figures are different.
	}
	\label{fig:Temp-ps-Sigma}
\end{figure*}

\begin{figure*}[t]
	\centering
	\includegraphics[width=\hsize,clip]{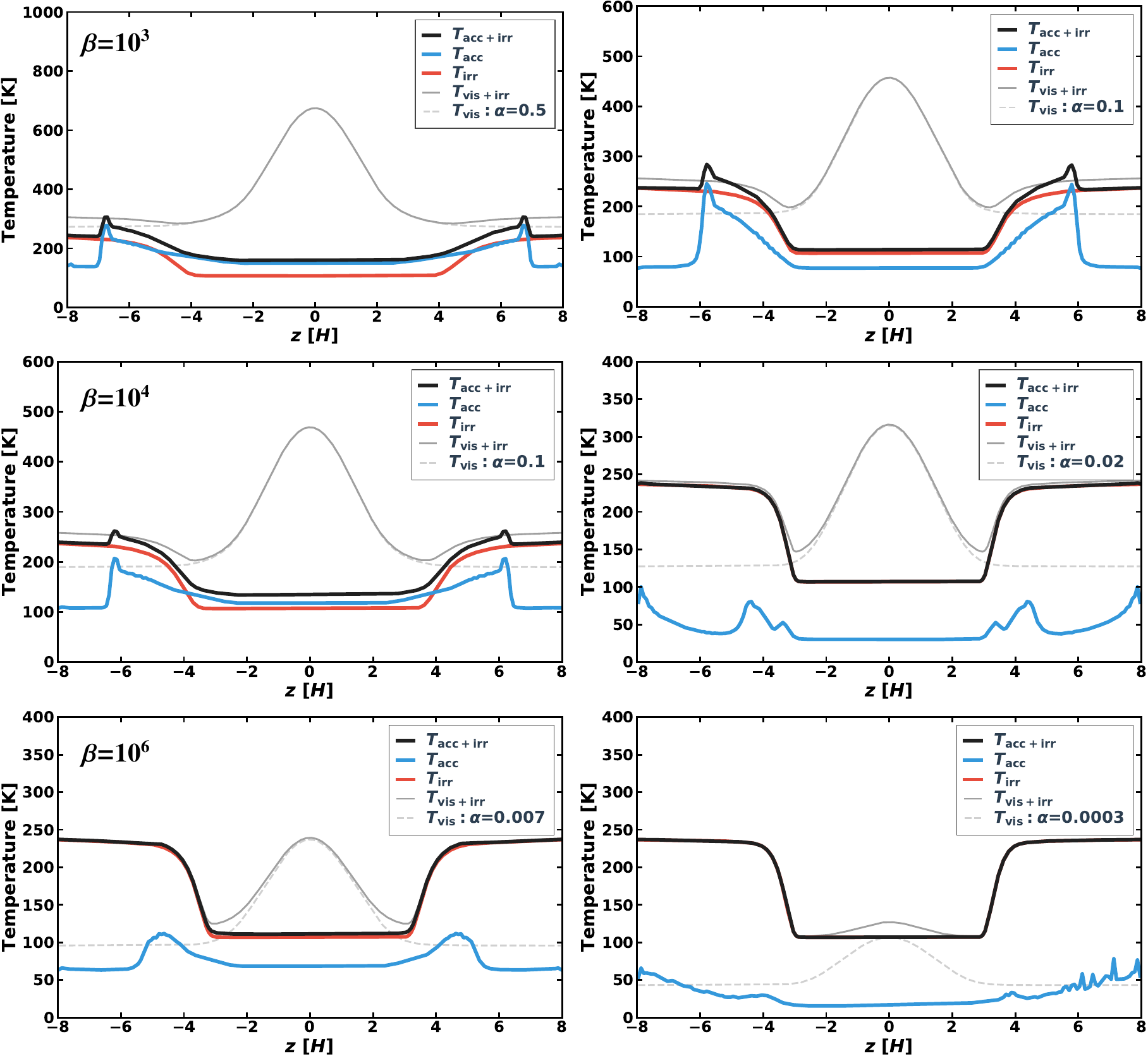}
	\caption{  	
	Same as Figure \ref{fig:Temp-ps-Sigma}, but for runs with $\beta=10^{3},10^{4},$ and $10^{6}$ (from top to bottom).
	}
	\label{fig:Temp-ps-beta}
\end{figure*}

\begin{figure*}[t]
	\centering
	\includegraphics[width=\hsize,clip]{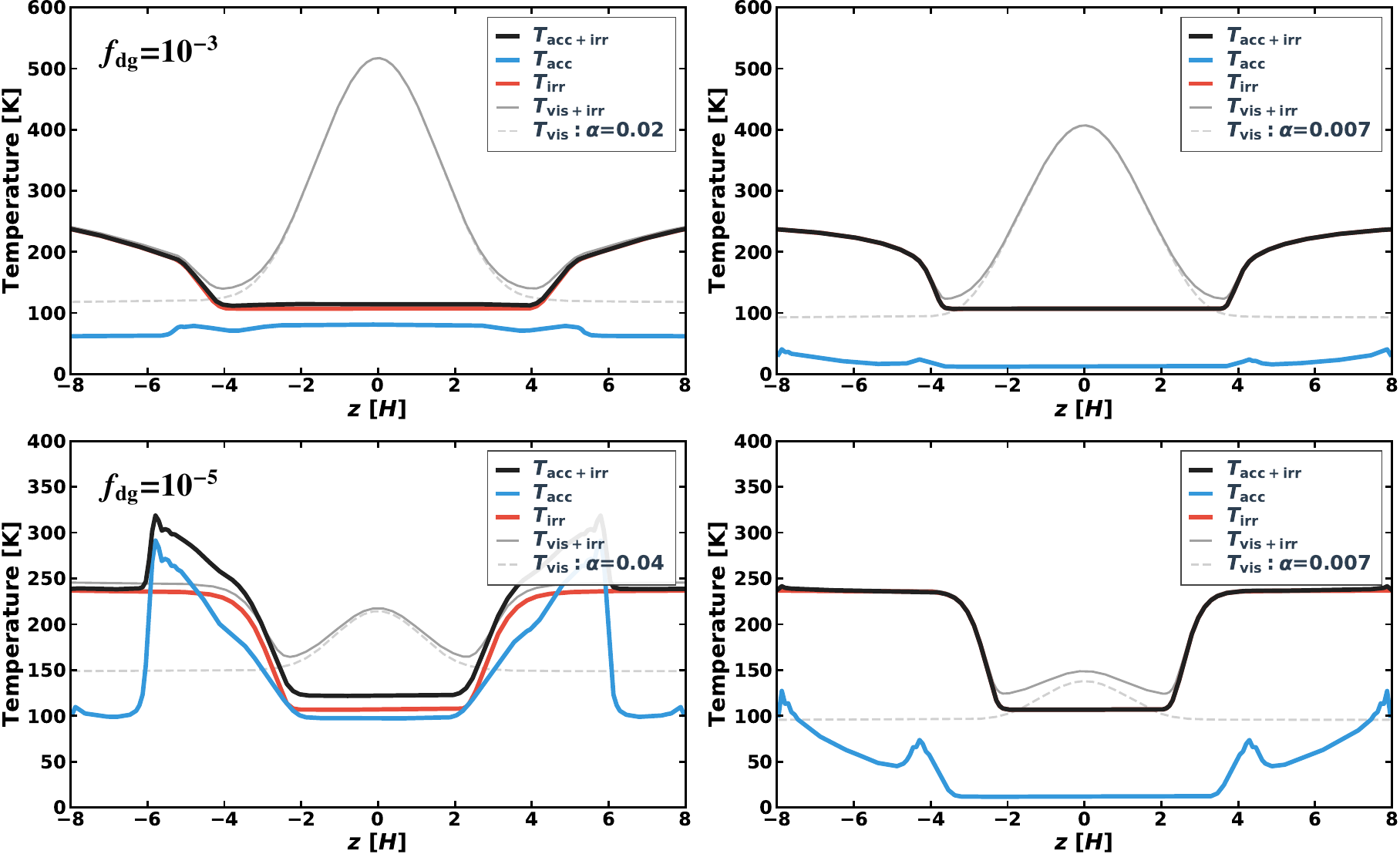}
	\caption{  	
	Same as Figure \ref{fig:Temp-ps-Sigma}, but for runs with $f_{\rm dg}=10^{-3},$ and $10^{-5}$ (from top to bottom).
	}
	\label{fig:Temp-ps-fdg}
\end{figure*}

To further access the role of accretion heating, we conduct a parameter study in this section, where
we vary the gas surface density $\Sigma$, the initial disk magnetization (characterized by $\beta_{0}$) and
the dust-to-gas ratio $f_{\rm dg}$. The results from varying $r$, distance to the star, is discussed separately in Section \ref{ssec:rad}. Both signs of
$B_{z}$ are considered in all cases. Compared with the fiducial simulations, we only vary one parameter at a time.
The range of parameters are described in Table \ref{tab:para},
and the results are summarized in Table \ref{tab:results}.

Figure \ref{fig:Temp-ps-Sigma} shows how the temperature profiles depend on the gas surface density. 
Increasing the surface density at fixed plasma $\beta$ gives higher magnetic field strength, higher mass accretion rate, associated with stronger heating. In the mean time, it gives higher optical depth, leading to more heat accumulation. On the other hand, the ionization level decreases with higher density, which reduces current and Joule dissipation.

For accretion heating from Joule dissipation, we find that while heating and the resulting temperature profile increases with increasing surface density, its overall contribution is still relatively small compared with irradiation unless the surface density is orders of magnitude higher. Also, heating from the $B_z>0$ case is much stronger than the $B_z<0$ case, where Joule dissipation is almost always negligible compared to irradiation. 
For heating from the constant-$\alpha$ viscous model, higher/lower accretion rate and gas surface density (i.e., higher/lower optical depth) both yield an increase/decrease of midplane temperature, leading to large/smaller differences compared with results from the Joule heating case.

In Figure \ref{fig:Temp-ps-beta}, we show results with different initial vertical magnetic field strength (characterized by $\beta_{0}$) in a way similar to 
Figure \ref{fig:Temp-ps-Sigma}. Obviously, stronger/weaker net vertical field gives higher/lower accretion rate (largely wind-driven).
We find that the variation of $\beta_0$ and does not strongly alter the location of Joule dissipation. 
Increasing/decreasing the field strength mainly enhances/reduces the total rate of Joule dissipation. 
The change in dissipation is more significant for the $B_z>0$ case, causing appreciable changes in midplane and surface temperatures, whereas in the $B_z<0$ case, disk temperature profile is again almost entirely determined by irradiation.
In the constant-$\alpha$ model, again, the midplane temperature is dominated by viscous dissipation and is sensitive to changes in accretion rate.

We then discuss the dependence of dust abundance in Figure \ref{fig:Temp-ps-fdg} by varying the dust-to-gas mass ratio of $10^{-3}$ and $10^{-5}$.
The dust abundance affects the ionization fraction and optical depth.
Higher dust abundance leads to higher the optical depth. It makes the optically thick region more extended (as seen in the $T_{\rm irr}$ profile), more heat accumulation, and hence higher midplane temperature. 
In the mean time, it leads to lower ionization fraction. This acts to suppress field growth, making dissipation take place at higher altitude, and hence reduce the contribution from Joule heating.
For lower dust abundance, lower optical depth tends to reduce midplane temperature, whereas the higher ionization fraction enhances Joule dissipation especially towards the surface (the disk remains laminar in our simulation), as well as its overall contribution to disk heating. We thus see prominent temperature bumps at disk surface in the $B_z>0$ case. For field polarities, lower grain abundance gives higher surface temperature.

\begin{figure*}[t]
	\centering
	\includegraphics[width=\hsize,clip]{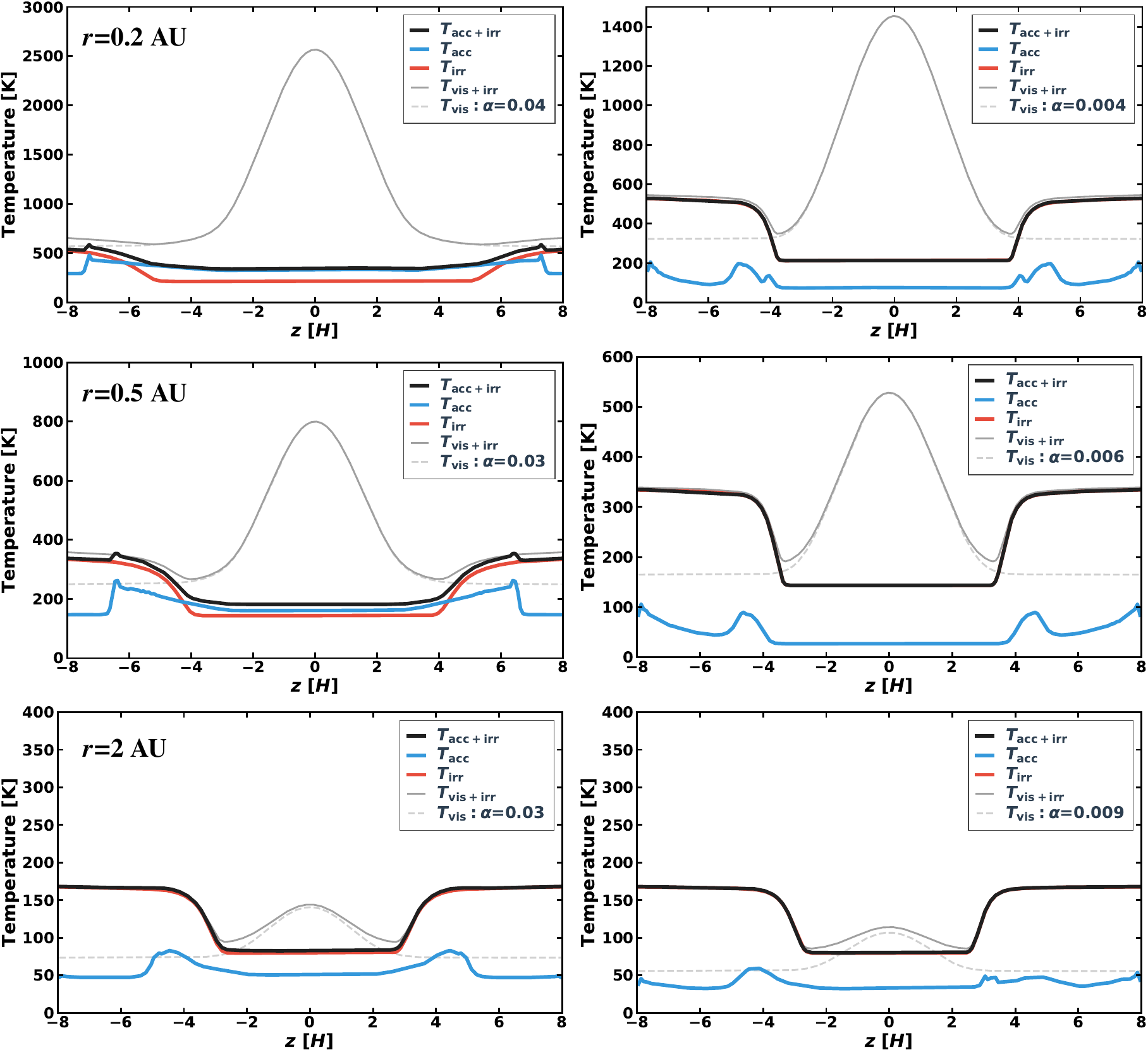}
	\caption{  	
	Same as Figure \ref{fig:Temp-ps-Sigma}, but for runs with $r=0.2,0.5,$ and $2$ AU (from top to bottom).	}
	\label{fig:Temp-ps-r}
\end{figure*}

\subsection[]{Dependence on Radial Distance}\label{ssec:rad}

In Figure \ref{fig:Temp-ps-r}, we discuss the dependence of accretion heating on the distance from the star, $r$. 
We first discuss the scenario from the conventional constant-$\alpha$ viscously-driven accretion model.
Towards larger distance, while irradiation gets weaker, it tends to play a more dominant role because viscous heating decreases with distance even faster. 
This is evident in our calculations assuming viscous heating with constant $\alpha$ parameters. Towards smaller $r$, we see that within $r\lesssim0.5$ AU, midplane temperature reaches and exceeds $\sim800$K in the constant-$\alpha$ model, which would trigger thermal ionization of Alkali species, and likely make the midplane region MRI active \citep{Desch2015High-temperatur}. This implies that MRI turbulence can be self-sustained in this region: viscous dissipation from the MRI maintains high midplane temperature needed for thermal ionization to sustain the MRI.

In the framework of our simulations (where disk temperature is fixed to irradiation temperature), the same trend holds in the case of $B_z>0$ in the sense that the role of Joule dissipation becomes more important towards smaller distances, and start to dominate over at $r\lesssim0.5$ AU. For the $B_z<0$ case, however, even at a close distance of $r=0.2$ AU, Joule dissipation is still negligible compared with irradiation heating. In both cases, the system temperature never gets close to $\sim800$K, the rough threshold for thermal ionization\footnote{Note that thermal ionization of alkali species is not included in our ionization chemistry model, and it is not needed as the results show.}, and in the simulations, the systems are well in the laminar state.
This means that the laminar states from our simulations are equally valid solutions, in addition to the case self-sustained MRI turbulence discussed earlier. 
Whether the system can stay in one case or the other then must be determined from global conditions and/or evolution history.

In reality, the innermost disk region (or more to the extreme, inner rim directly illuminated by the star) is likely warm enough to be MRI turbulent, which must be separated from a cooler laminar region somewhere further out. This interface has been conventionally designated as the inner dead zone boundary. Previous numerical studies already found it to be highly dynamic \citep[e.g.,][]{Faure2014Thermodynamics-,Flock20173D-Radiation-No}, though only Ohmic resistivity was considered and angular momentum transport/Joule heating from a disk wind was absent. Our result further suggests that this boundary should also be accompanied with abrupt temperature transitions, which may further complicate its dynamics, with potentially important implications on planet formation in this region.


\section{Discussion}\label{sec:discuss}

\subsection{Geometry of Magnetic Field}\label{ssec:disc-sym}

We have employed local shearing-box simulations to study the vertical distribution of current density.
One important issue of this approach is that a horizontal current layer that accompanies a flip of the sign of the
horizontal magnetic field (which is necessary for a physical field geometry in a disk wind) tends to be unstable in shearing-box 
simulations as pointed out by \citet{Bai2013aWind-driven-Acc}.
They found that the natural geometry of global magnetic field in a shearing box is such 
that the field lines have no flip and the horizontal fields on the top and bottom sides 
of the box have the same direction.  
This tendency is also observed in our simulations: even though we start with magnetic field geometry  
of a flip at the midplane, the flip gradually moves toward high altitude and eventually escapes
from the simulation box through the vertical boundary. 
The instability of the horizontal current layer seems to occur because 
the field lines straighten out under magnetic tension that is amplified 
by the Keplerian shear and the Hall effect.

In reality, the magnetic field threading a protoplanetary disk should have global geometry 
such that the field lines are directing outward on both side of the disk, and hence the horizontal field
should have a flip at some height within the disk. 
Depending on its height, the current layer accompanied by this flip could contribute 
to the heating of the disk interior, but this cannot be evaluated in our local simulations.

Recent global simulations by \citet{Bai2017Global-Simulati} that include all three non-ideal MHD effects found that the flip occur naturally in global simulations. In particular, in their fiducial model with parameters similar to ours, thanks to the Hall effect, the flip occurs at very high altitude (about 4-5$H$ above the midplane) on one side of the disk in the inner disk (a few AU). The location of the flip roughly coincides with the location of the FUV front. Because of such high altitude, magnetic field profiles in the bulk disk below the location of the flip are in fact similar to the profiles obtained in shearing-box. Therefore, on the one hand, our calculations miss additional heating resulting from the strong current layer due to the flip. On the other hand, this single-sided heating at very high altitude likely only causes very localized heating near the disk atmosphere (see Figure \ref{fig:a-wo-sym} in Appendix \ref{sec:app} for an example), and has very limited impact to the disk midplane temperature.
Meanwhile, we plan to address this issue further with global simulations in future works.

\subsection{Dependence on the Prior Temperature}\label{ssec:disc-temp}

The approach we have taken to study disk heating is not self-consistent: although the temperature profiles obtained using \eqref{eq:T} properly take into account Joule heating and irradiation heating, the Joule heating rates are obtained from isothermal MHD simulations. 
For this reason, the ``posterior'' temperature obtained from \eqref{eq:T} differs from the ``prior'' temperature given in the MHD simulations.
Although self-consistent modeling will be the subject of our future work, it is important to clarify within the current approach how much a variation of the prior temperature can influence the resulting heating rate 
and posterior temperature. 

\begin{figure*}[t]
	\centering
	\includegraphics[width=0.49\hsize,clip]{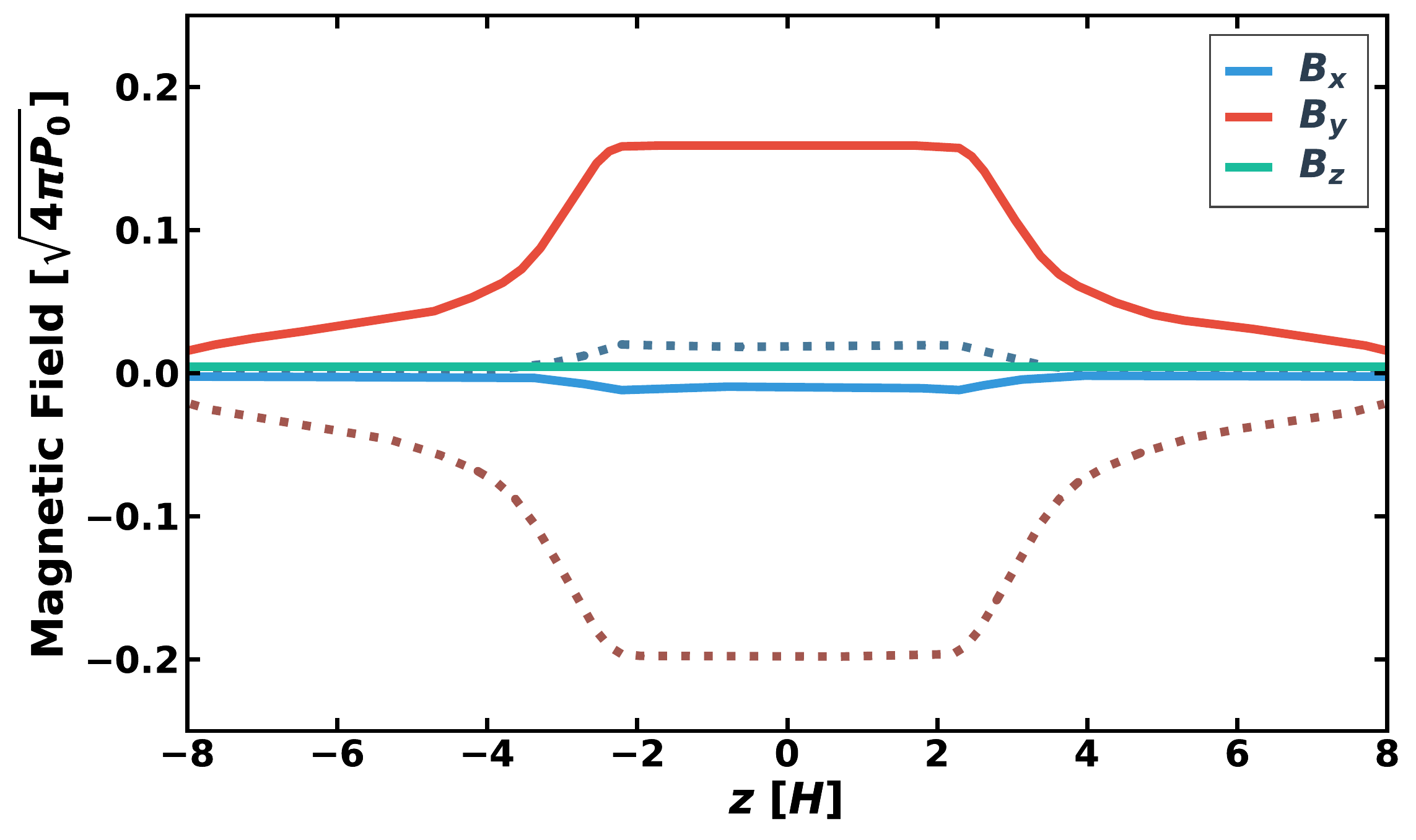}
	\includegraphics[width=0.49\hsize,clip]{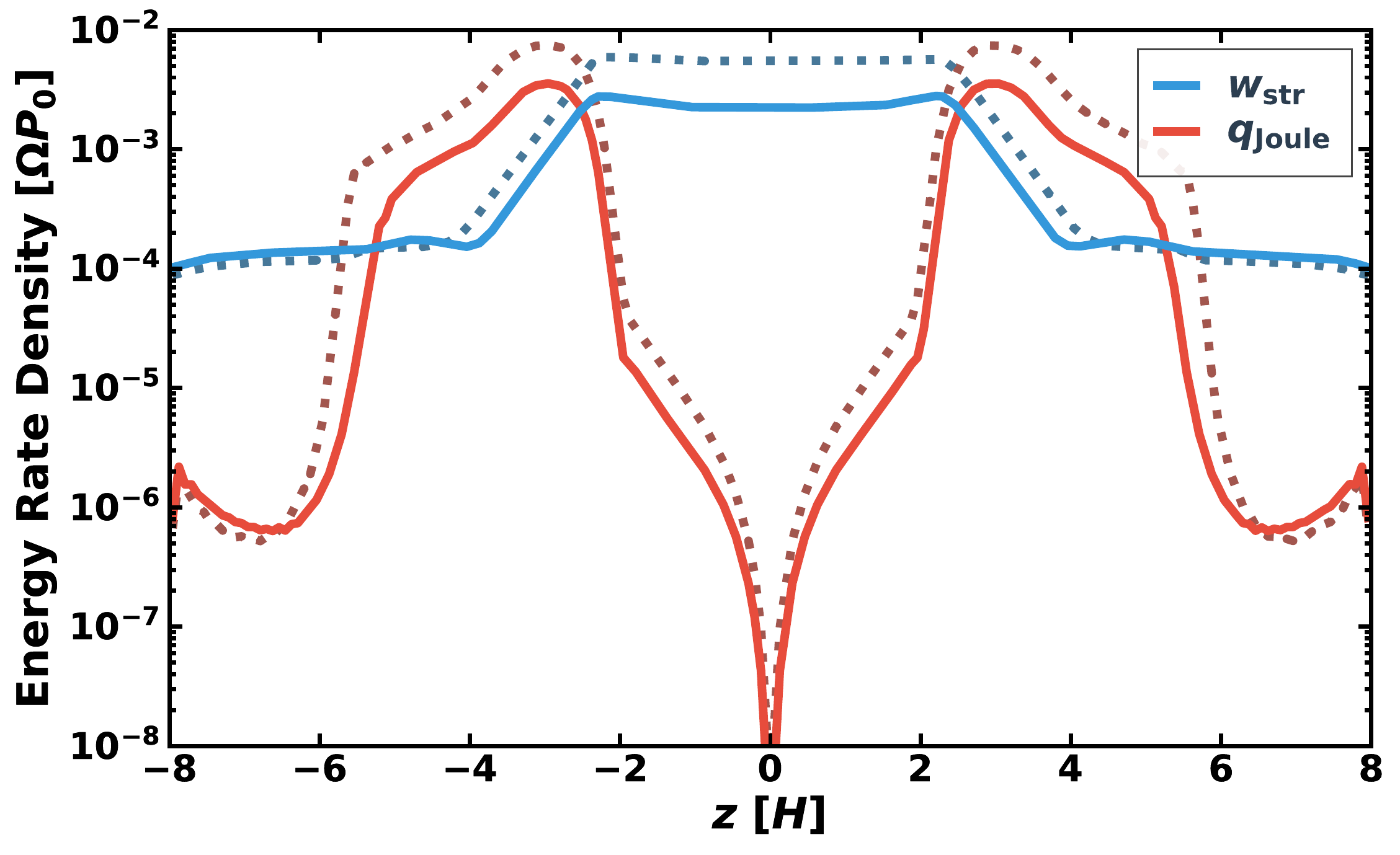}
	\caption{ 
	Same as the top panels of Figure \ref{fig:fid+}, 
but for the run with $T=280$ K.
	Dotted lines show the results of the fiducial run for comparison.
	}
	\label{fig:dep-temp}
\end{figure*}

For this purpose, we repeated the fiducial run
but with a prior temperature of $T=280$ K,
which is approximately twice the fiducial value.
This choice is motivated by the fact that 
for passively irradiated disks,
the temperature in the optically thin surface region 
is about two times higher than that in 
the optically thick disk interior 
\citep[e.g.,][]{Chiang1997Spectral-Energy}.
Figure \ref{fig:dep-temp} compares the vertical profiles 
of the stress and accretion heating rates from the fiducial runs with the two different prescribed temperatures.\footnote{
In this simulation run, the signs of the horizontal fields $B_x$ and $B_y$ are opposite to those in the original fiducial run. However, this has no physical significance because the equations governing the local shearing box are invariant under the transformation $(x,y) \to (-x,-y)$. The polarity of the horizontal field in the saturation state is randomly determined depending on the initial perturbations. 
} 
It shows that in normalized units, the heating rate in this case reduced by a factor of two.
This difference mainly comes from the temperature dependence on the diffusivities.
Since the posterior temperature given by \eqref{eq:T} scales with the heating rate as weakly as $q^{1/4}$, 
a variation of the prior temperature only weakly
affects the posterior temperature.

\subsection{Impacts of Inefficient Accretion Heating on Planet Formation}

Conventionally, the temperature profiles of PPDs in planet formation studies are obtained by adopting a simple viscous accretion disk model, similar to our constant-$\alpha$ model. In the more realistic situation of wind-driven accretion with a largely laminar disk, our results indicate that accretion heating of the disk interior is much less efficient, and viscous heating models could overestimate the temperature near the midplane, where planet formation mainly proceeds. These results could have a number of implications for planet formation.

For example, one important constraint can be derived on the formation history of
the solar-system rocky planets including the Earth \citep{Oka2011Evolution-of-Sn,Sato2016On-the-water-de,Morbidelli2016Fossilized-cond}. 
The fact that the water content of the solar-system terrestrial planets 
is tiny implies that they formed interior to the snow line. 
Constant-$\alpha$ disk models infer that the inner region of the solar nebula where the terrestrial planet formed retained a temperature above the sublimation point of water ice as long as the nebular accretion rate 
is comparable to or above the median accretion rate of classical T-Tauri stars, $10^{-8}~M_\sun~\rm yr^{-1}$ \citep{Davis2005Condensation-Fr,Oka2011Evolution-of-Sn,Bitsch2015The-structure-o}. 
However, in the absence of viscous heating, the nebular temperature 
at heliocentric radii of $\approx 1~\rm au$ 
must have fallen below the ice sublimation point
as the young Sun's luminosity decreased to the present-day solar 
luminosity \citep{Kusaka1970Growth-of-Solid,Chiang1997Spectral-Energy,Turner2012A-Hot-Gap-aroun}
, which likely occurred in $\approx$ 1 Myr after Sun's formation \citep{Turner2012A-Hot-Gap-aroun,Bitsch2015The-structure-o}.
This would imply that the solar-system rocky planets either formed very early 
($\la 1 ~\rm Myr$), or had formed closer to the sun and subsequently 
migrated outward to arrive at their present-day positions.
The latter scenario is consistent with a recent model 
of rocky planet formation invoking nebular gas dispersal due to disk winds \citep{Ogihara2017Effects-of-glob}.

Another important implication is related to the fact that the inner disk region $R\lesssim0.5$ AU may possess either a cold laminar state without thermal-ionization driven by disk winds, or a hot MRI-turbulent state with thermal ionization (see our earlier discussions on Figure \ref{fig:Temp-ps-r}). This fact suggests that the dynamics of such innermost disk regions are complex and may exhibit state transitions depending on the history of evolution. Such complex behaviors are already hinted from MHD simulations that include different levels of thermodynamics/radiative transfer and Ohmic resistivity \citep{Faure2014Thermodynamics-,Faure2015Vortex-cycles-a,Flock20173D-Radiation-No}, and may have profound implications to planet formation \citep[e.g.,][]{Chatterjee2014Inside-out-Plan}. In addition, the rate and direction of type-I planet migration are known to sensitively depend on the thermodynamic structure of PPDs \citep{Tanaka2002Three-Dimension,Bitsch2015The-structure-o}, again requiring reliable understanding of the disk heating mechanisms that we have studied.

\subsection{On Plasma Heating by Strong Electric Fields}

We have neglected change of the ionization fraction due to strong electric fields \citep{Inutsuka2005Self-sustained-,Okuzumi2015The-Nonlinear-O}.
In the case of MRI turbulence, the electric field may heat up electrons, enhance its adsorption onto grains, reduce the ionization fraction, and in turn further damp the MRI \citep{Okuzumi2015The-Nonlinear-O,Mori2016Electron-Heatin,Mori2017Electron-Heatin}. Much stronger electric field, on the other hand, may trigger electric discharge (known as lightning), and thereby increase of the ionization level that promotes the MRI \citep{Inutsuka2005Self-sustained-,Muranushi2009Direct-Simulati}.
While our simulations are laminar, the layer especially where horizontal magnetic field flips may possess substantial current density, that might make the system enter this regime of non-linear Ohm's law. This effect will be addressed in future publications. 

\section{Summary and Conclusion}\label{sec:Summary}

In this work, we have investigated the temperature profiles in the inner region of PPDs, where recent studies have suggested that the weakly ionized disks are largely laminar with accretion primarily driven by magnetized disk winds. Correspondingly, accretion heating mostly takes the form of Joule dissipation instead of viscous dissipation. To this end, we have performed quasi-1D shearing-box non-ideal MHD simulations to quantify the Joule dissipation profiles, based on which disk vertical temperature profiles are calculated.
We start from analyzing accretion heating with fiducial parameters, followed by a parameter exploration. The results are summarized as follows.
\begin{itemize}

\item 
The energy dissipation due to Joule heating in PPDs is the strongest at relatively high altitudes ($z\sim3H$), as a result of poor conductivity at disk midplane. This leads to little heat accumulation in the midplane region, and hence reduced midplane temperature.

\item 
The Joule heating profile depends on the polarity of net vertical magnetic fields threading the disk (even though the wind stress does not), due to the Hall effect. It is enhanced in the aligned ($\bm{B}_0\cdot\bm{\Omega}>0$) case due to the Hall-shear instability, and is strongly reduced in the anti-aligned case.

\item
At a given accretion rate, Joule heating is much less efficient than viscous heating,
yielding much smaller midplane temperatures especially in the inner disk regions.
Varying disk surface density, radial location, magnetization and dust abundances only weakly affect the conclusions above.

\item
As long as the disk remains largely laminar, Joule dissipation only plays a minor-to-modest (aligned) or even negligible (anti-aligned) role compared to stellar irradiation in determining disk temperature profiles, in standard disk models. However, an MRI-turbulent state can also be sustained in the very inner disk ($\lesssim0.5$ AU) where viscous dissipation raises disk temperature to trigger thermal ionization.

\end{itemize}

This study shows that accretion heating in the wind-driven accretion PPDs is much weaker than commonly assumed. It also highlights the importance of stellar irradiation rather than the accretion heating in determining PPD temperatures even in the early stages of disk evolution. More self-consistent simulations in full three dimensions are needed to better address the coupling between radiative processes and gas dynamics, which requires coupling non-ideal MHD with radiative transfer as well as non-thermal and thermal ionization physics. Meanwhile, consequences of these results to planet formation remain to be explored. 

\acknowledgments 
The authors thank Hubert Klahr, Oliver Glessel, Neal Turner, Hidekazu Tanaka, Shigenobu Hirose, Kengo Tomida, Kazunari Iwasaki, and Takahiro Ueda for fruitful discussion and meaningful comments.
The authors also thank the anonymous referee for comments that improved the paper. 
This work was supported by JSPS KAKENHI Grant Number JP15H02065, JP16K17661, JP16H04081, and JP17J10129. XNB acknowledges support from Tsinghua University.
Numerical computations were carried out on Cray XC30 and XC50 at Center for Computational Astrophysics, National Astronomical Observatory of Japan.

\appendix

\section{Temperature structure of reflection-asymmetric dissipation profile in disks}\label{sec:app}

We here derive the analytic expressions of the temperature profile with a general dissipation profile
by solving the radiative transfer equation of thermal radiation.

We extend the derivation of \citet{Hubeny1990Vertical-struct} by taking into account the following two effects.
\citet{Hubeny1990Vertical-struct} assumed that the work done by the viscous stress is locally dissipated in a
Keplerian disk. However, this is no longer true in the wind-driven scenario, as we see in Figure \ref{fig:fid+},
which we now take into account.
In addition, the stellar irradiation offers another heating source.
Here, we use the rate profile $q(z)$, being the sum of dissipation and irradiation heating,
as the energy source term
\begin{equation}
	q =  q_{\rm Joule} + q_{\rm irr} \ .
\end{equation}
We assume that the incoming radiation from irradiation and outgoing radiation by dust thermal emission
are in visible and infrared, respectively.
This approach allows us to separately solve the radiation fields \citep{Calvet1991Irradiation-of-}, and
here we only consider radiative transfer of radiation reemitted by dust.
The second assumption in \citet{Hubeny1990Vertical-struct} is that energy source term has reflection
symmetry across the midplane.
However, this no longer holds when energy dissipation occurs at one side of the disk, since the flip
of the toroidal magnetic field generally occurs at one side of the disk, as seen in global non-ideal MHD
simulations \citep[e.g.,][]{Bai2017Global-Simulati}.
Some of our simulations also show similar asymmetric structures.

We solve the radiative transfer equations of zeroth, first, and second moments of specific intensity $I(z,\mu,\nu)$ of the cosine of the incident angle $\mu$.
We define these moments as
\begin{eqnarray}
	\vvec{  J_{\nu} \\  H_{\nu} \\  K_{\nu} }  \equiv  \frac{1}{2} \int_{-1}^{1}  I(z,\mu,\nu) \vvec{ 1 \\ \mu \\  \mu^{2} } \dmu \ .
\end{eqnarray}
In addition, we also define the frequency-integrated moments with its frequency $\nu$ in the thermal wavelength, respectively:
\begin{eqnarray}
	\vvec{  \Jth \\  \Hth \\  \Kth }  \equiv   \int_{\rm thermal}   \vvec{ J_{\nu} \\  H_{\nu} \\  K_{\nu} } \d\nu \ .
\end{eqnarray}
The zeroth and first moments of the radiative transfer equation integrated over thermal wavelength are written as, respectively,  
\begin{eqnarray}
	\pd{\Hth}{z} &=& \rho \kappa_{B_{\rm th} } B_{\rm P} - \rho \kappa_{J{\rm th}} \Jth \ , \label{eq:a-Hz} \\
	\pd{\Kth}{z} &=& - \rho \kappa_{H{\rm th}} \Hth  \ ,  \label{eq:a-Kz}
\end{eqnarray}
where $\kappa_{J{\rm th}}$, $\kappa_{B{\rm th}}$, and $\kappa_{H{\rm th}}$ are the absorption mean opacity, Planck mean opacity, and flux mean effective opacity, respectively \citep{Mihalas1978Stellar-atmosph}:
\begin{eqnarray}\label{eq:a-kapJ}
	\kappa_{J{\rm th}} &=& \Jth^{-1} \int_{\rm thermal} \frac{ \alpha_{\nu} }{  \rho} J_{\nu} \d\nu \ , \\
	\kappa_{B{\rm th}} &=&  B_{\rm P}^{-1} \int_{\rm thermal} \frac{ \alpha_{\nu}  }{  \rho} B_{\nu} \d\nu \ , \\
	\kappa_{H{\rm th}} &=& \Hth^{-1} \int_{\rm thermal} \frac{ \alpha_{\nu} + \sigma_{\nu} }{  \rho} H_{\nu} \d\nu \ ,
\end{eqnarray}
where $\alpha_{\nu}$ and $\sigma_{\nu}$ are the coefficients of true absorption and scattering respectively,
and $B_{\rm P}=\sigma T^{4} /\pi$ is the frequency-integrated Planck function.
Here, we assume that $\kappa_{J{\rm th}} = \kappa_{B{\rm th}} $, and both of them are equal to the Rosseland
mean opacity $\kappa_{\rm R}$. 

The second basic equation is the energy balance between the energy absorption and the thermal radiation.
The vertical gradient of the energy flux $\mathcal{F} = 4 \pi \Hth $ of radiative transport is equal to the
energy dissipation rate per unit volume.
We also neglect the energy transport due to gas motion (e.g., advection and convection) for simplicity,
which holds when such timescales are long compared to the timescale to establish thermodynamic equilibrium.
Energy conservation is then expressed as
\begin{equation}\label{eq:a-Q}
	 4\pi \pd{ \Hth }{z} = q \ .
\end{equation}

To close the radiative transfer equations, we adopt the Eddington approximation, which assumes isotropic
radiation field \citep[e.g.,][]{Mihalas1978Stellar-atmosph,Rybicki1979Radiative-proce}, and it gives the
relation
\begin{eqnarray}
	\frac{\Kth(z)}{\Jth(z)} &=& \frac{1}{3} \ .       \label{eq:a-kj-ed} 
\end{eqnarray}
In addition, for outgoing boundary conditions, we adopt the two stream approximation, where the outgoing radiation is characterized by
\begin{eqnarray}
	\frac{\Kth(+\infty)}{\Jth(+\infty)} &=& \frac{\Kth(-\infty)}{\Jth(-\infty)}  = \frac{1}{3} \ ,       \label{eq:a-kj} \\
	\frac{\Hth(+\infty)}{\Jth(+\infty)} &=& -\frac{\Hth(-\infty)}{\Jth(-\infty)} = \frac{1}{ \sqrt{3} }  \ ,  \label{eq:a-hj}  
\end{eqnarray}
which are valid for the optically thick regions.

Using above assumptions, we integrate Equations (\ref{eq:a-Q}) and (\ref{eq:a-Kz}) from $z$ to
$+\infty$ to obtain
\begin{equation}\label{eq:a-H}
	\Hth(z) = \Hth(+\infty) - \int_{z}^{+\infty}\frac{q}{4 \pi} \dz' \ ,
\end{equation}
\begin{equation}\label{eq:a-K}
	\Kth(z) = \Kth(+\infty) + \int_{z}^{+\infty} \rho \kappa_{H{\rm th}} \Hth  \dz' \ .
\end{equation}
From these equations and the boundary conditions, we can calculate $\Hth(z)$ and $\Kth(z)$.
Using Equations (\ref{eq:a-kj-ed}), (\ref{eq:a-kj}), and (\ref{eq:a-K}), we obtain
\begin{equation}\label{eq:a-J}
	\Jth(z) = \Jth(+\infty)  +   3 \int_{z}^{+\infty} \rho \kappa_{H{\rm th}} \Hth  \dz'  \ .
\end{equation}
Using Equations (\ref{eq:a-Hz}), (\ref{eq:a-Q}), and (\ref{eq:a-J}), the temperature profile is expressed as 
\begin{eqnarray}\label{eq:a-T0}
	 \frac{\sigma T^{4}(z)}{\pi}  
	 &=&     \pr{  3 \int_{z}^{+\infty} \rho \kappa_{H{\rm th}} \Hth  \dz'  +  \Jth(+\infty) } 
	 	+ \frac{1}{\rho \kappa_{\rm R} }\frac{q}{4 \pi} \ ,
\end{eqnarray}
where we have used $B=\sigma T^{4}/\pi$. Using Equation \eqref{eq:a-hj}, the temperature
profile is further expressed as
\begin{eqnarray}\label{eq:a-T1}
	 T(z) =  \pr{\frac{4 \pi \Hth(+\infty) }{\sigma}}^{1/4} \pr{  
	  \frac{3}{4 \Hth(+\infty)}  \int_{z}^{+\infty} \rho \kappa_{H{\rm th}} \Hth  \dz'  
	 + \frac{\sqrt{3}}{4}  
	 +\frac{1}{4 \Hth(+\infty)\rho \kappa_{\rm R}} \frac{q}{4 \pi}  }^{1/4}\ .
\end{eqnarray} 

If $\Hth(+\infty)$ is known, with $\Hth$ given by \eqref{eq:a-H}, the temperature profile can
be directly obtained by \eqref{eq:a-T1}. Using \eqref{eq:a-H}, we see
\begin{equation}\label{eq:a-Hinf-}
	\Hth(+\infty) - \Hth(-\infty) = \frac{\Gamma}{4 \pi} \ ,
\end{equation}
where
\begin{equation}
		\Gamma = \int_{-\infty}^{+\infty} q \dz \ 
\end{equation}
is the total heating (dissipation and irradiation) rate. When the dissipation profile is
symmetric, we have $H_{\rm th}(+\infty)=-H_{\rm th}(-\infty)=\Gamma/8\pi$ and the
derivation is complete.
However, this does not necessarily hold without the reflection symmetry. In the general case,
we substitute \eqref{eq:a-H} into \eqref{eq:a-K}, and $\Kth(z)$ can be written as 
\begin{equation}\label{eq:a-K2}
	\Kth(z) = \Kth(+\infty) +  \Hth(+\infty) \int_{z}^{+\infty} \rho \kappa_{H{\rm th}} \dz' -  \int_{z}^{+\infty} \rho \kappa_{H{\rm th}} \pr{ \int_{z'}^{+\infty}\frac{q}{4 \pi} \dz'' } \dz'   \ .
\end{equation}
Taking $z = -\infty$ in \eqref{eq:a-K2},  we find $H(+\infty)$ to be
\begin{equation}\label{eq:a-Hinf1}
	\Hth(+\infty) =  \frac{1}{ \tau_{H, \rm tot}  }\pr{ - \Delta K_{+\infty}   +    \int_{-\infty}^{+\infty} \rho \kappa_{H{\rm th}} \pr{  \int_{z'}^{+\infty} \frac{q}{4 \pi}\dz''  }  \dz'  } \ ,
\end{equation}
where
\begin{equation}
	\Delta K_{\infty} = \Kth(+\infty) - \Kth(-\infty) \ ,
\end{equation}
\begin{equation}
	\tau_{H, \rm  tot} = \int_{-\infty}^{+\infty} \rho \kappa_{H{\rm th}} \dz \ .
\end{equation}
If $\Delta K_{+\infty}$ is given, we can then obtain $\Hth(+\infty)$. 
Using Equations (\ref{eq:a-kj}) and  (\ref{eq:a-hj}), the sum of $H(+\infty)$ and $H(-\infty)$ is expressed as,
\begin{equation}\label{eq:a-Hinf+}
	\Hth(+\infty) + \Hth(-\infty) = \sqrt{3}\Delta K_{\infty} \ .
\end{equation}
Combining Equations (\ref{eq:a-Hinf+}) and (\ref{eq:a-Hinf-}), we eliminate 
$\Hth(-\infty)$ and obtain
\begin{equation}\label{eq:a-dK}
	\Delta K _{\infty} =  \frac{1}{\sqrt{3}} \pr{ 2\Hth(+\infty) - \frac{\Gamma}{4 \pi} }   \ .
\end{equation}
Substituting \eqref{eq:a-dK} into \eqref{eq:a-Hinf1}, we obtain
\begin{eqnarray}\label{eq:a-Hinf}
	\Hth(+\infty) =  \frac{1}{  \tau_{H, \rm tot} + 2/\sqrt{3}   }\pr{ \frac{\Gamma}{4\sqrt{3}\pi}  +    \int_{-\infty}^{+\infty} \rho \kappa_{H{\rm th}} \pr{  \int_{z}^{+\infty} \frac{q}{4 \pi} \dz'  }  \dz  }\ .
\end{eqnarray}
This is the general expression for $H(+\infty)$ that allows for asymmetric heating profiles.

\begin{figure}[t] 
	\centering
	\includegraphics[width=0.49\hsize,clip]{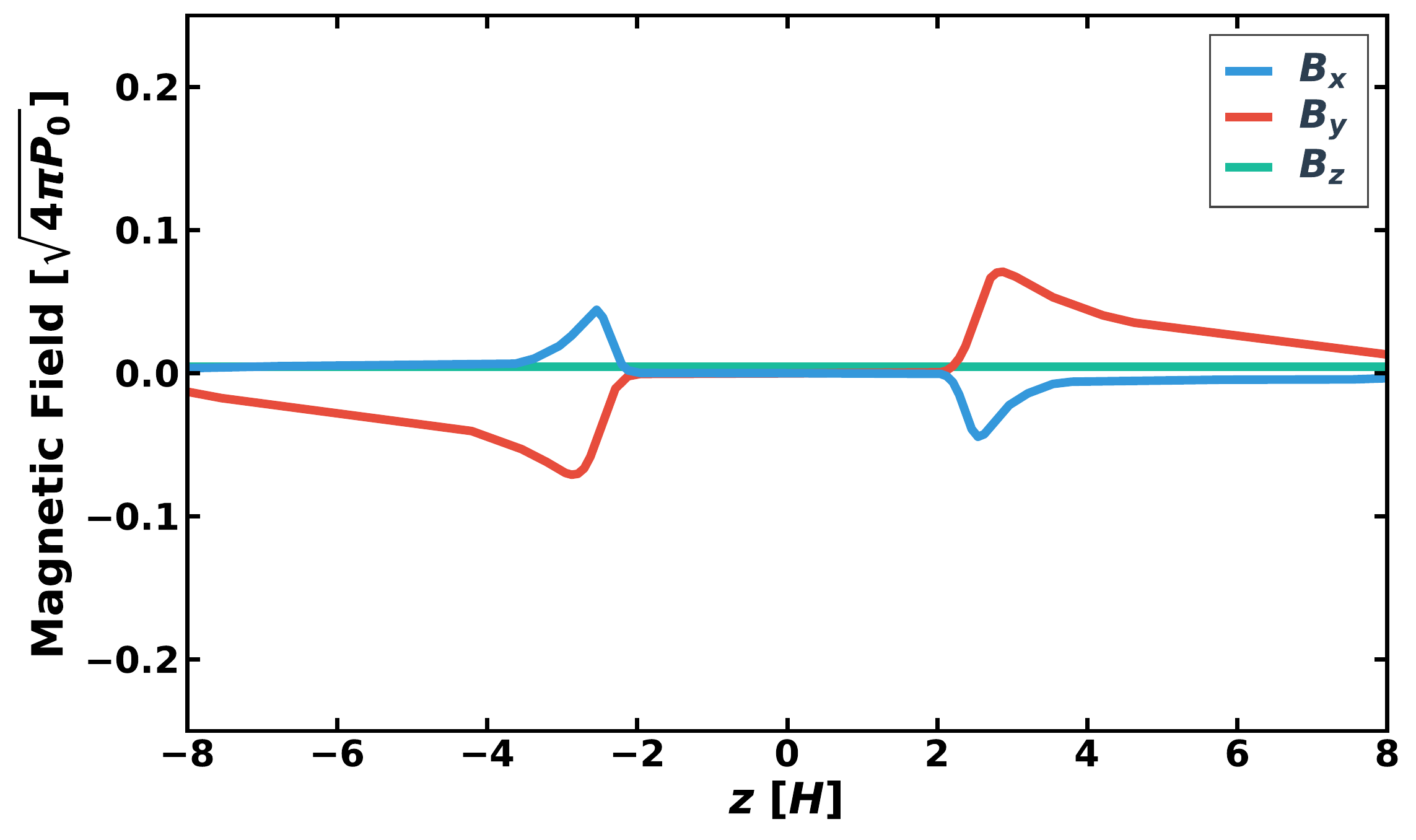}	
	\includegraphics[width=0.49\hsize,clip]{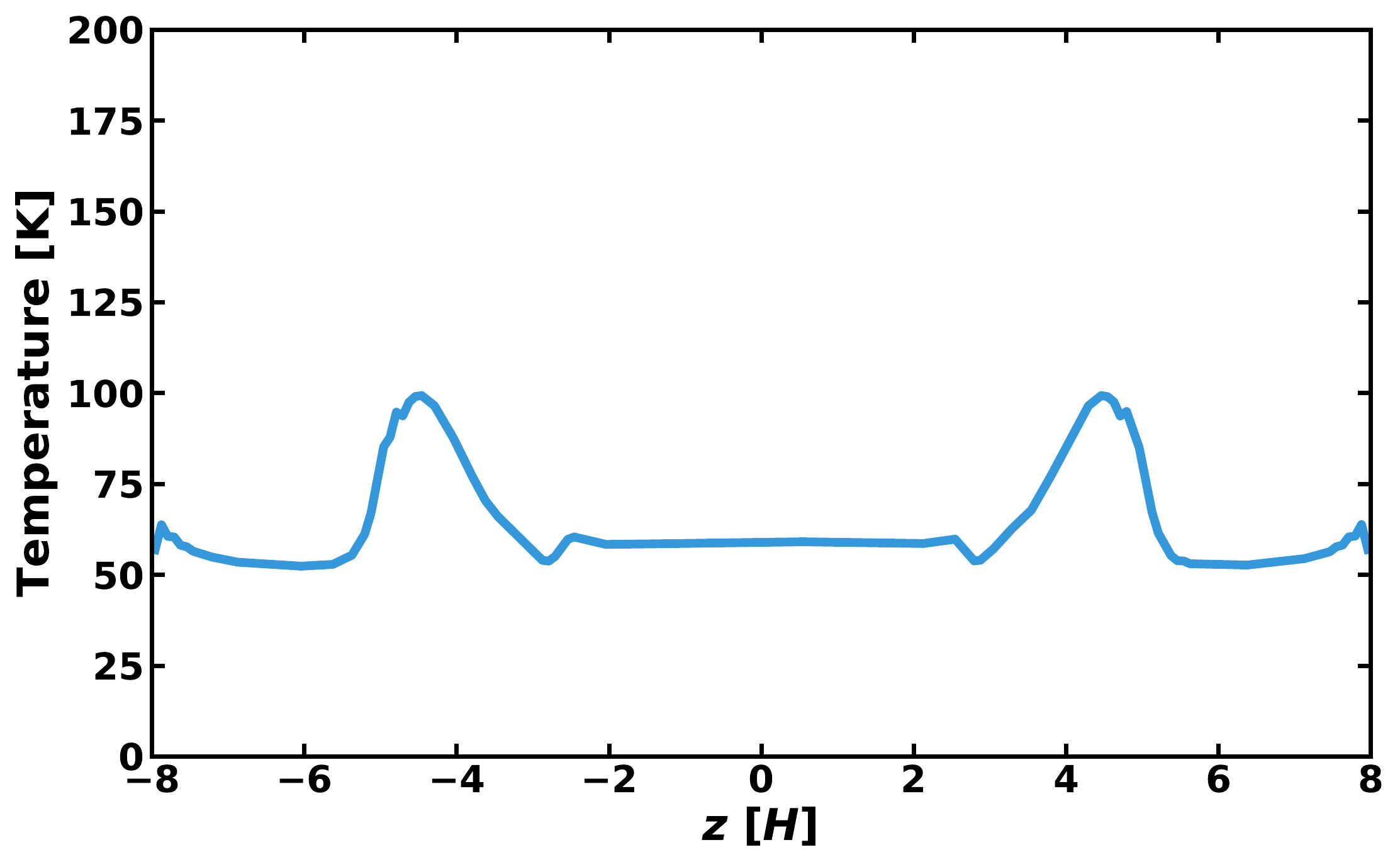}\\
	\includegraphics[width=0.49\hsize,clip]{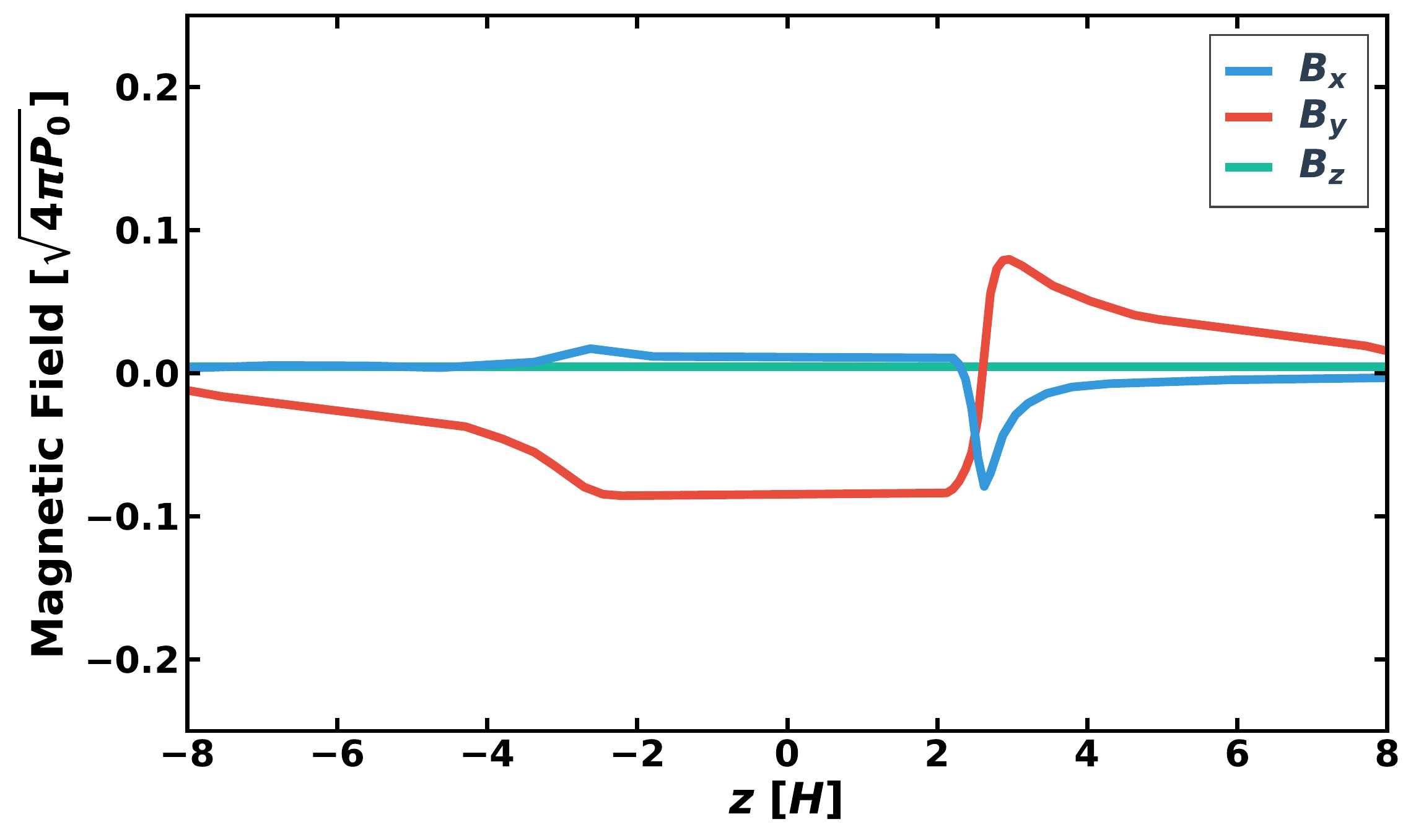}	
	\includegraphics[width=0.49\hsize,clip]{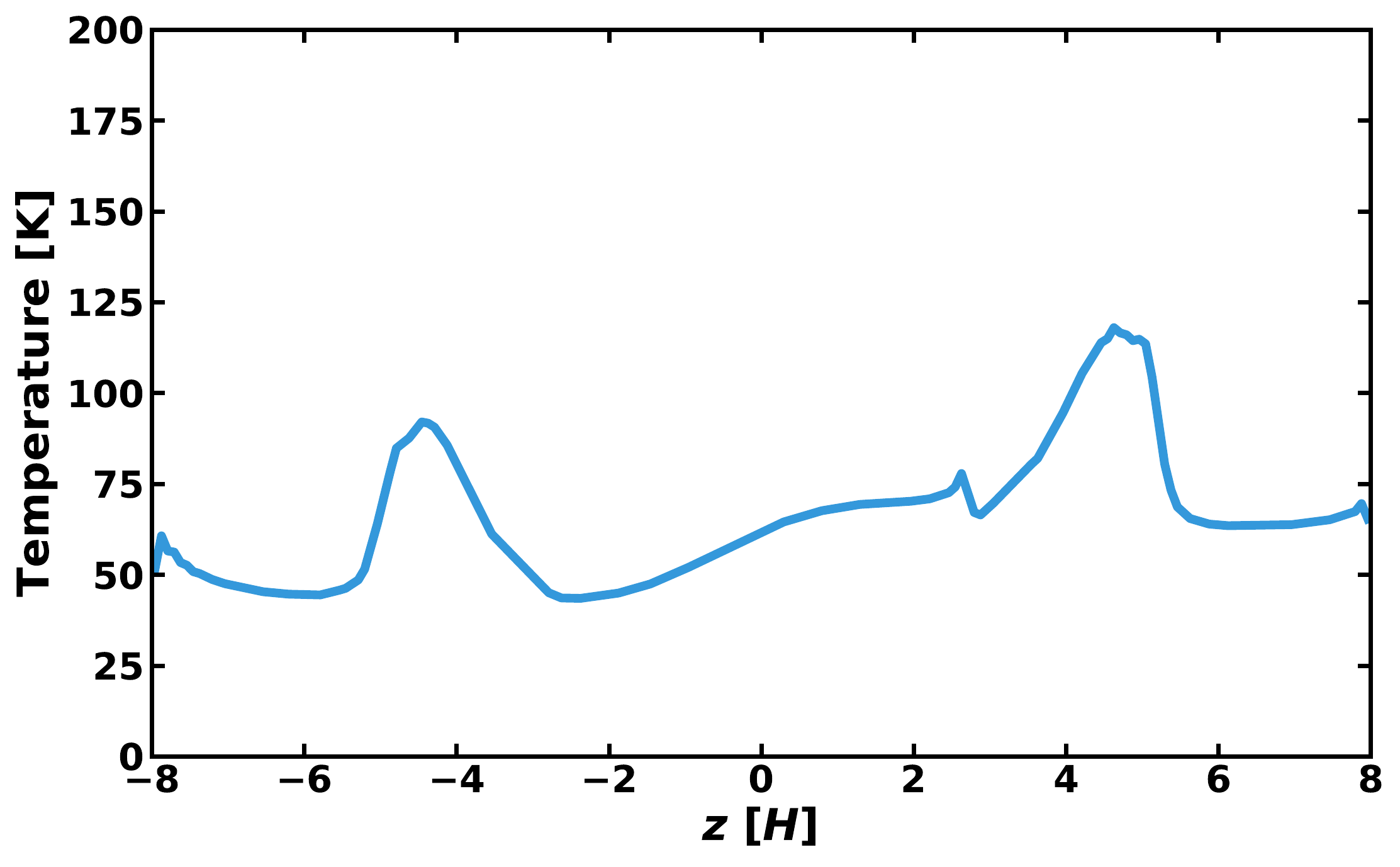}\\
	\includegraphics[width=0.49\hsize,clip]{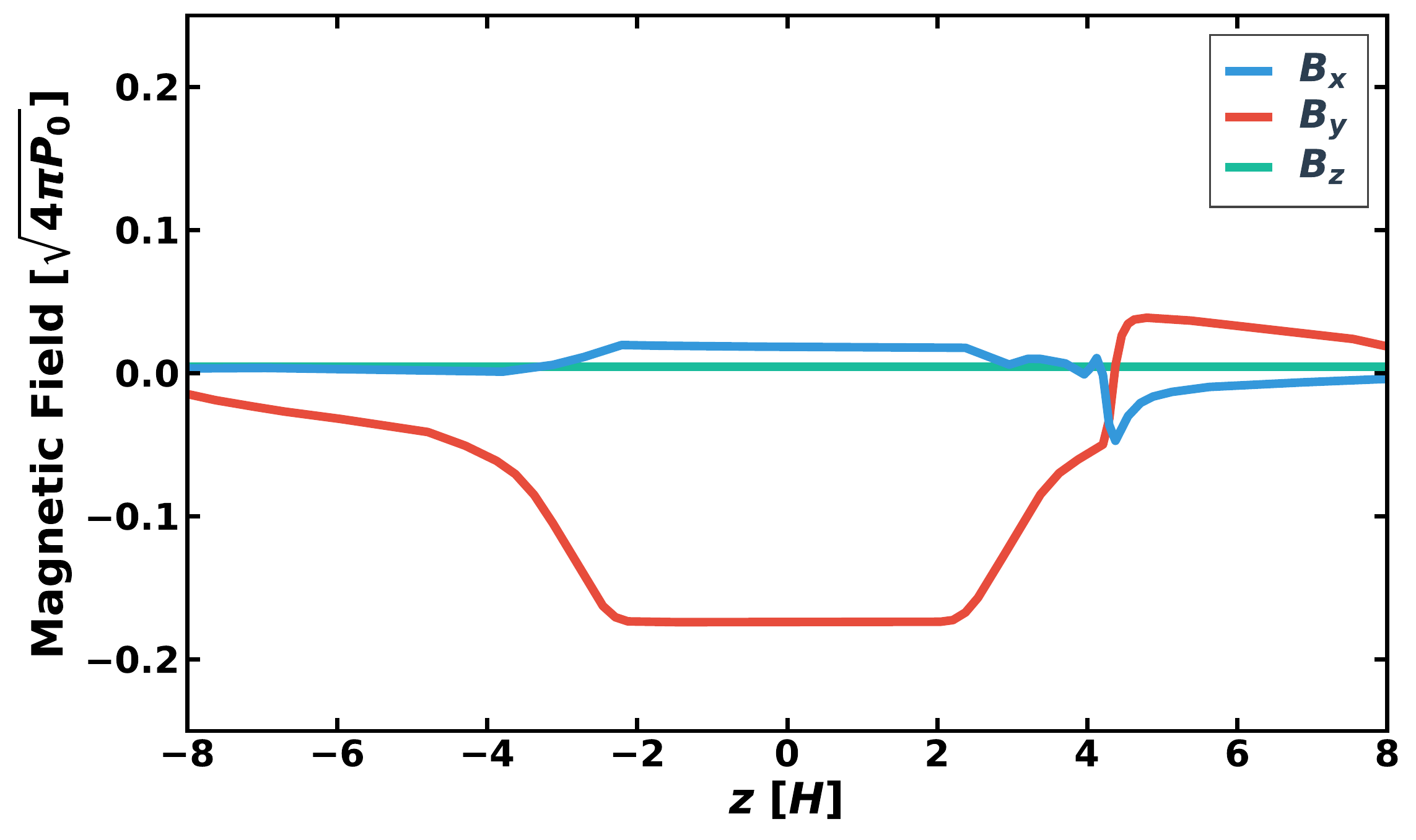}	
	\includegraphics[width=0.49\hsize,clip]{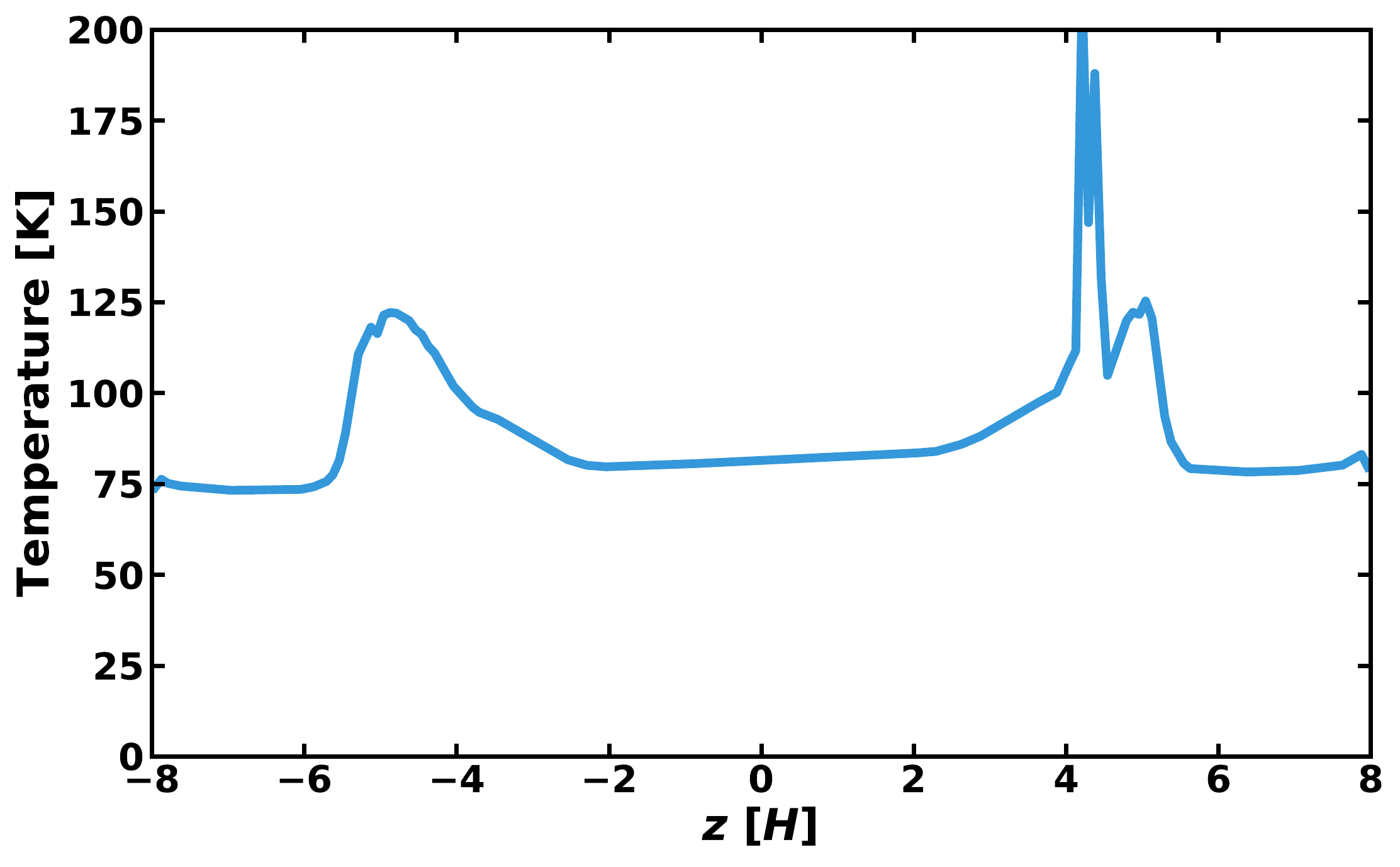}				
	\caption{  	
	Snapshots of vertical profile of magnetic field (left) and temperature determined only by Joule heating (right) for the fiducial run at relatively early time of $4.9$, $12.7$, $13.9$ orbits.
	The physical magnetic field geometry is sustained until $\sim 12$ orbits.
	This field geometry is not preserved at later time as horizontal field flips to become symmetric about the midplane (due to limitation of local simulations).
	}
	\label{fig:a-wo-sym}
\end{figure}

Knowing $H(+\infty)$, we finally derive the temperature profile.
Using Equations (\ref{eq:a-T1}) and (\ref{eq:a-Hinf}), the temperature profile is derived as 
\begin{equation}\label{eq:a-Teq}
	 T(z)  =   T_{\rm eff}          
	 \pr{     
	     \frac{ 3}{4}  \tau_{\rm eff}(z)  
	 +   \frac{  \sqrt{3}  }{ 4  }    	 
	 +    \frac{q}{ 4 \rho \kappa_{\rm R} \mathcal{F}_{+\infty}}
	 }^{1/4} \ ,
\end{equation}
where we use the radiative flux at the upper surface
$\mathcal{F}_{+\infty} = 4 \pi \Hth(+\infty)$, and
$T_{\rm eff} =  (\mathcal{F}_{+\infty} / \sigma )^{1/4}$ is the effective temperature observed from the upper side, and
\begin{equation}
	\tau_{\rm eff}(z) = \int_{z}^{+\infty} \rho\kappa_{\rm R}  \pr{1 - \frac{1}{\mathcal{F}_{+\infty} }\int_{z'}^{+\infty} q \dz'' } \dz'  \ , \label{eq:a-taueff-2}
\end{equation}
is the effective optical depth.
Here, we further take $\kappa_{H{\rm th}} = \kappa_{B{\rm th}}=\kappa_R$ by assuming that the
scattering coefficient is much smaller than the absorption coefficient, $\sigma_{\nu}/\alpha_{\nu}\ll 1$.
The first term in \eqref{eq:a-Teq} expresses the effect that heat accumulation increases disk temperature.
When the dissipation profile is reflection symmetric, the temperature profile reduces to that
described in \citet{Hubeny1990Vertical-struct}.

We note that the definition of effective optical depth, tracking back to its original expression in \eqref{eq:a-T1}, represents a
radiative-flux-weighted optical depth
\begin{equation}
	\tau_{\rm eff}(z) = \frac{1}{\mathcal{F}_{+\infty}} \int_{z}^{+\infty} \rho\kappa_{\rm R}  \mathcal{F}(z) \dz'  \ . \label{eq:a-taueff}
\end{equation}
To better understand its physical meaning, we consider two extreme cases.
When the all accretion energy is released at the midplane, above the midplane the radiative flux is equal to the outgoing flux $\mathcal{F}_{+\infty}$, and hence this value is equal to the standard optical depth $\tau_{R, \rm tot}/2$.
In the second case, assume all accretion energy is released at a height $z=\pm z_{\rm heat}$.
In the region between the heating positions above and below the midplane, the radiative flux is zero because the flux from the upper and lower sides cancels out. Correspondingly, $\tau_{\rm eff}(z)$ becomes
\begin{equation}
	\tau_{\rm eff}(z) =  \left\{
	\begin{array}{ll}
		 \displaystyle\tau_{\rm col}(z)  &, (|z| \geqq  z_{\rm heat}) ,\\
		\displaystyle \tau_{\rm col}(z_{\rm heat})  &, (0 \leqq |z| < z_{\rm heat}),
	\end{array} \right.
\end{equation}
where
\begin{equation}
	\tau_{\rm col} (z) = \displaystyle\int_{z}^{+\infty} \rho \kappa_{\rm R} \dz \, .
\end{equation}
Especially, the effective optical depth $\tau_{\rm eff}(z=0)$ at the midplane is simply the optical depth at $z_{\rm heat}$, which can be much smaller than the actual midplane optical depth due to the weighting. Also
note that the effective optical depth can even become negative at the midplane if large energy dissipation occurs at high altitude. 

Finally, we consider an example with asymmetric heating profile. In doing so, we show in Figure \ref{fig:a-wo-sym} snapshots of the magnetic field profiles and the temperature profile resulting from only accretion heating together for the fiducial run with $B_{z}>0$ at an early evolution time of 5, 12.7, and 13.9 orbits.
In this case, horizontal magnetic field flips at one side leaving a current sheet with strong
dissipation.\footnote{Similar to the case discussed in \citet{Bai2015Hall-Effect-Con}, the system first evolves into an asymmetric profile with horizontal field flipping at a height offset from the midplane by several scale heights. This asymmetry is not long-lived owing to the limitation of local simulation, whereas it can be preserved in global simulations (\citealp{Bai2017Global-Simulati}, see their Figure 7, and further discussions in Section 5.1).}
We see that higher temperatures at the lower side than at the upper side, with a spike
at the current sheet, with additional temperature peaks at the disk surface where vertical
gradient of $B_y$ is strong.

\section{Conservation of Mechanical Energy in the Simulations}\label{sec:cons}

\begin{figure}[t] 
	\centering
	\includegraphics[width=0.5\hsize,clip]{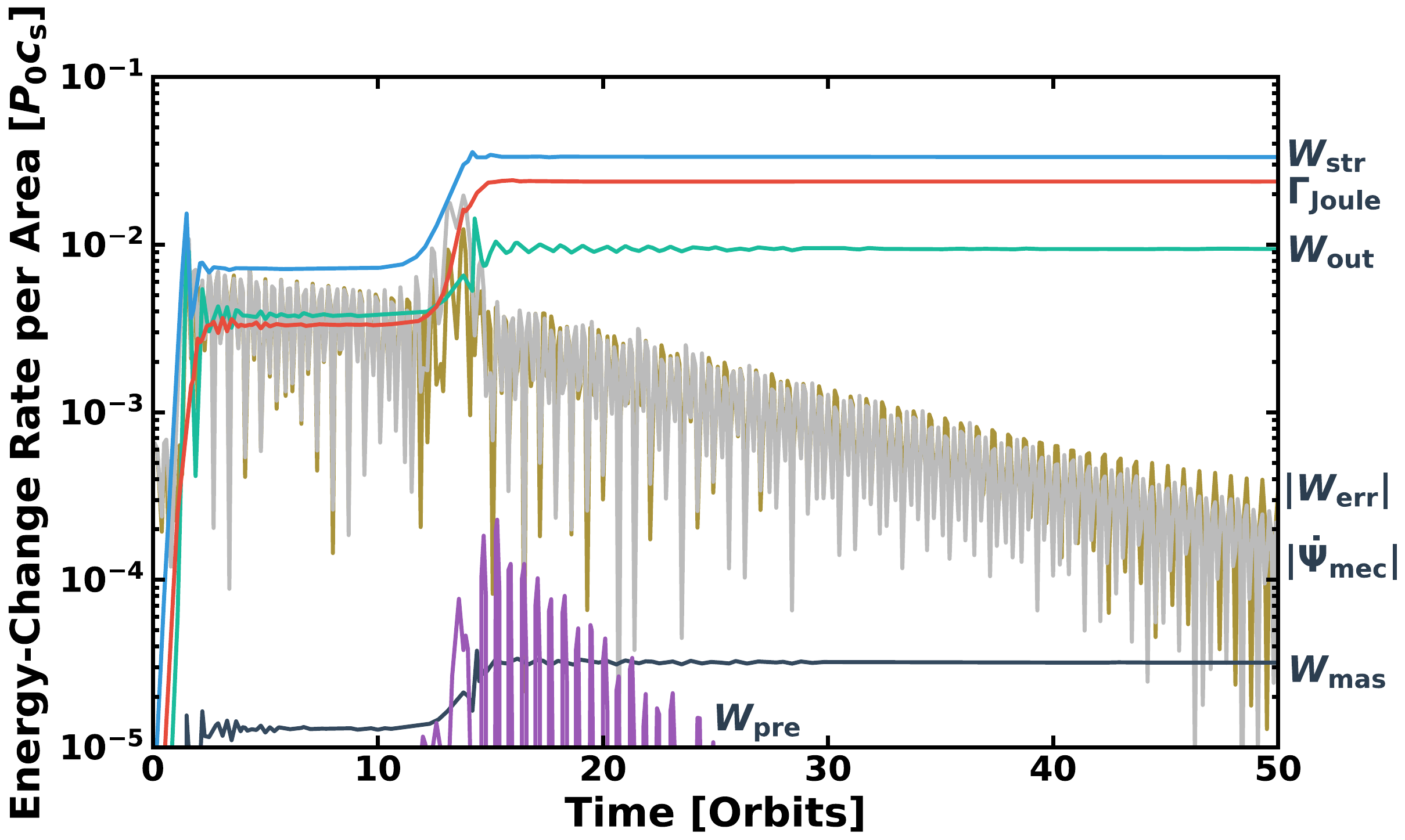}	
	\caption{Result from the fiducial run with $B_z>0$. Shown are the
	time variation of the time derivative of the total energy $\dot{\Psi}_{\rm mec}$, 
	the rate of work done by the Reynolds and Maxwell stress $W_{\rm str}$, 
	the rate of energy dissipation due to resistivity $\Gamma_{\rm Joule}$, 
	the rate of energy outflow thorough vertical boundary $W_{\rm out}$,
	 the rate of work done on the fluid by pressure $W_{\rm pre}$, 
	 the rate of additional energy due to the mass conservation $W_{\rm mas}$,
	 and the rest of the energy rates $W_{\rm err}$ in the simulation domain, which are described in Equations (\ref{eq:whole-energy-rate})--(\ref{eq:W-press}). 
	}
	\label{fig:t-E}
\end{figure}

To demonstrate energy conservation in our simulations, we first integrate
\eqref{eq:mech-energy-rate} over the computational domain to
obtain the rate of change in the total mechanical energy per area as
\begin{equation}\label{eq:whole-energy-rate}
	  \dot{\Psi}_{\rm mec} = \frac{1}{L_{x}L_{y}}\pd{}{t} \int E_{\rm mec} \dV
	  =  W_{\rm str}  - W_{\rm out}  -  \Gamma_{\rm Joule} + W_{\rm pre} ,
\end{equation}
where 
\begin{equation}
	W_{\rm out} =  \frac{1}{L_{x}L_{y}}\pr{\int_{z=L_{z}/2} -  \int_{z=-L_{z}/2}} F_{\rm mec} \dx\dy \, 
\end{equation}
is the rate of energy loss through the vertical boundary (by disk winds),
\begin{equation}\label{eq:Winj}
	W_{\rm str} =   \frac{3 \Omega}{2L_{y}}   \int_{x=L_{x}/2} \pr{\rho v_{x} \delta v_{y} - \frac{B_{x}B_{y}}{4 \pi} } \dy\dz \, 
\end{equation}
is the rate of energy injection through the Reynolds and Maxwell stresses, 
\begin{equation}\label{eq:GamJoule}
	\Gamma_{\rm Joule} =   \frac{1}{L_{x}L_{y}} \frac{1}{c}\int \bm{J} \cdot \bm{E}' \dV \, 
\end{equation}
is the rate of energy dissipation due to Joule heating, and 
\begin{equation}\label{eq:W-press}
	W_{\rm pre} =  \frac{1}{L_{x}L_{y}}  \int P \nabla \cdot \bm{v}  \dV  \, 
\end{equation}
is the work done on the fluid by pressure per unit time.
The integrate of the second term in the left-hand-side of \eqref{eq:mech-energy-rate} is
eliminated because the term is a periodic quantity in the $y$-direction.
In addition to these energy rates, we consider the other energy rates.
The energy rate due to mass added for mass conservation is described by $W_{\rm mas}$.
We also define the rest of the energy rates as
$W_{\rm err} = \dot{\Psi}_{\rm mec} - (W_{\rm str} - W_{\rm out} - \Gamma_{\rm Joule} + W_{\rm pre} + W_{\rm mas})$.

Taking the simulation with $B_z>0$, we show in Figure \ref{fig:t-E}
the time evolution of each term in
the equation for the mechanical energy \eqref{eq:whole-energy-rate}.
We see that despite small oscillations (presumably due to breathing mode that is
leftover from initial evolution), $W_{\rm err}$ and $\dot{\Psi}_{\rm mec}$
diminishes in time, and the system converges into a steady state which is fully
laminar. It also implies time average should be performed at late times, which
we choose to be between 40 and 50 orbits.
Over this period, we find that about 71.4\% and 28.3\% of $W_{\rm str}$ is used for the energy
dissipation of Joule heating and the energy outflow by the disk wind, respectively, indicating excellent level of mechanical energy conservation (note that while Athena conserves {\it total energy}, there can be small truncation errors in mechanical energy conservation).

\bibliography{mybib} 

\end{document}